\documentclass[]{llncs}
\usepackage{graphicx} 
\usepackage{subfigure}
\usepackage{xspace}
\usepackage{todonotes}
\usepackage{caption}
\usepackage{comment}
\usepackage{amsfonts}
\usepackage{colortbl}
\usepackage{hyperref}
\usepackage[inline]{enumitem}
\usepackage{paralist}
\usepackage{hhline}

\PassOptionsToPackage{table,xcdraw}{xcolor}

\definecolor{myblue}{RGB}{4,178,230}
\definecolor{myorange}{RGB}{252,111,45}

\newcommand{\myblue}[1]{\textcolor{myblue}{#1}}
\newcommand{\myorange}[1]{\textcolor{myorange}{#1}}

\newcommand{\ped}{\textsc{ped}\xspace}
\newcommand{\peds}{\textsc{ped}s\xspace}
\newcommand{\med}{\textsc{med}\xspace}
\newcommand{\meds}{\textsc{med}s\xspace}
\newcommand{\smed}{\textsc{smed}\xspace}

\newcommand{\hmed}{\textsc{hmed}\xspace}

\newcommand{\shmed}{\textsc{shmed}\xspace}
\newcommand{\shmeds}{\textsc{shmed}s\xspace}

\graphicspath{{figs/}}

\begin{document}

\title{
Evaluating Animation Parameters\\for Morphing Edge Drawings\thanks{Research partially supported by: $(i)$ University of Perugia, Ricerca di Base 2021, Proj. ``AIDMIX — Artificial Intelligence for
Decision Making: Methods for Interpretability and eXplainability''; $(ii)$ MUR PRIN Proj. 2022TS4Y3N - ``EXPAND: scalable algorithms for EXPloratory Analyses of heterogeneous and dynamic Networked Data''; $(iii)$ MUR PRIN Proj. 2022ME9Z78 - ``NextGRAAL: Next-generation algorithms for constrained GRAph visuALization''; $(iv)$ University of Perugia, Ricerca di Base, grant RICBA22CB; $(v)$ DFG grant Ka 812-18/2.}
}

\author{Carla Binucci\inst{1}\orcidID{0000-0002-5320-9110} \and
Henry Förster\inst{2}\orcidID{0000-0002-1441-4189} \and
Julia Katheder\inst{2}\orcidID{0000-0002-7545-0730} \and
Alessandra Tappini\inst{2}\orcidID{0000-0001-9192-2067}}
\institute{University of Perugia, Perugia, Italy \email{\{carla.binucci,alessandra.tappini\}@unipg.it}
\and
University of Tübingen, Tübingen, Germany 
\email{\{henry.foerster,julia.katheder\}@uni-tuebingen.de}
}

\date{}

\maketitle

\begin{abstract}
Partial edge drawings (\ped) of graphs avoid edge crossings by subdividing each edge into three parts and representing only its stubs, i.e., the parts incident to the end-nodes.
The morphing edge drawing model (\med) extends the \ped drawing style by animations that smoothly morph each edge between its representation as stubs and the one as a fully drawn segment while avoiding new crossings.
Participants of a previous study on \med (Misue and Akasaka, GD19) reported eye straining caused by the animation. We conducted a user study to evaluate how this effect is influenced by varying animation speed and animation dynamic by considering an easing technique that is commonly used in web design.
Our results provide indications that the easing technique may help users in executing topology-based
tasks accurately. The participants also expressed appreciation for the easing and a preference for a slow animation speed.

\end{abstract}

\section{Introduction}
Edge crossings are well-known to reduce the readability and the perceived aesthetics of graph drawings; see, e.g.,~\cite{DBLP:conf/gd/Purchase97,DBLP:journals/ese/PurchaseCA02}. Bruckdorfer and Kaufmann~\cite{DBLP:conf/fun/BruckdorferK12} suggested a quite rigorous solution that avoids crossings in straight-line drawings by drawing edges only partially. More precisely, in a \emph{partial edge drawing} (\ped) of a graph,
each edge is subdivided into three parts and the middle part is not drawn; the drawn parts are called \emph{stubs}.
While the model has received some attention in follow-up studies (see, e.g.,~\cite{DBLP:conf/iisa/BinucciLMT16,DBLP:journals/jgaa/BruckdorferCGKM17,DBLP:conf/gd/Bruckdorfer0L15,DBLP:journals/jgaa/BruckdorferKM14,DBLP:conf/iv/Burch17,DBLP:conf/gd/BurchVKW11,DBLP:conf/gd/HummelKNN19,DBLP:conf/grapp/SchmauderBW15}), its practical applicability seems to be hindered by the fact that \peds cannot be read as quickly as traditional node-link diagrams~\cite{DBLP:conf/gd/Bruckdorfer0L15}. On the other hand, there are indications that they can be interpreted more accurately than drawings where edges are fully drawn~\cite{DBLP:conf/gd/Bruckdorfer0L15}.

Recently, Misue and Akasaka~\cite{DBLP:conf/gd/MisueA19} suggested to enhance the \ped drawing style by a sequence of animations that smoothly morph each edge between its partial representation and its representation as a fully drawn segment.
With the resulting drawing style, known as \emph{morphing edge drawing} (\med), 
%
the goal is to help the users read drawings more quickly while maintaining the readability offered by \ped. A user study by Misue and Akasaka~\cite{DBLP:conf/gd/MisueA19} indicates that the model can achieve these goals.

An unfortunate side effect of the addition of the 
animation appears to be eye fatigue experienced by the users: Namely, while the users of Misue's and Akasaka's study~\cite{DBLP:conf/gd/MisueA19} reported that  in fact ``[morphing] made it easy to confirm the exact adjacency'', they also stated that ``[their] eyes [were] strained.'' The users also indicated potential reasons for this negative effect stating that ``[it was] messy and difficult to [focus on]'', that ``the stubs [changed] too fast'', and that ``the time for stubs to connect [was] too short.'' The latter two aspects have been considered in Misue's follow-up study~\cite{DBLP:conf/gd/Misue22}, where the speed of the morphing animation was substantially reduced while there was a small delay added to the animation  at the time that the two stubs meet. 
It is worth noting that -- aside from these two changes -- both previous studies on \med~\cite{DBLP:conf/gd/Misue22,DBLP:conf/gd/MisueA19} have mainly focused on minimizing the total animation duration, which is proposed to be a main factor in user response time. However, the existing user study's focus~\cite{DBLP:conf/gd/MisueA19} was not to investigate different animation speeds, but to establish the validity of \med in comparison to \ped and traditional straight-line drawings.

We also remark that \meds may be regarded as \emph{staggered} animations between \peds and traditional straight-line drawings. These types of animations have been previously investigated for morphs between two 2D scatterplots of the same data points~\cite{DBLP:journals/tvcg/ChevalierDF14}. Curiously, in contrast to Misue and Akasaka's findings for \med~\cite{DBLP:conf/gd/MisueA19}, Chevalier et al.~\cite{DBLP:journals/tvcg/ChevalierDF14} did not observe indications for benefits of staggered morphing animations between two scatterplots.

\paragraph{Our contribution.} We study \meds from a user-centric point of view. To this end, we investigate in a user study to which extent animation speed influences user response time and accuracy in the execution of tasks. Moreover, to counteract the eye straining reported by the participants of the previous user study~\cite{DBLP:conf/gd/MisueA19}, we 
smoothen the animation by an approach called \emph{easing} that is widely used in webdesign~\cite{mdn,DBLP:conf/cgi/IzdebskiKS20} and experimentally evaluate its impact\footnote{Research data and all stimuli available at \url{https://github.com/tuna-pizza/MEDSupplementaryMaterial}.}. We remark that to the best of our knowledge, easing has not been evaluated in the context of graph animations before. Our results indicate that fast animation and easing can help users in performing topology-based tasks. Also, while participants appreciated the easing technique, they preferred a slower speed.

\smallskip
\noindent The paper is structured as follows. Section~\ref{se:med} introduces \meds and the animation parameters we consider. Section~\ref{se:user-study} describes the design of our user study. Section~\ref{se:results} discusses the quantitative and the subjective results of our experiment, as well as its limitations. Section~\ref{se:conclusion} lists some future research directions.

\section{Morphing Edge Drawings}\label{se:med}

\paragraph{Model Description.} A \med of a graph $G=(V,E)$ is a \emph{dynamic} drawing, i.e., a function $\Gamma: G \times T \rightarrow \mathbb{R}^2$ where $T$ is the time interval of the drawing animation. In this paper, we study straight-line \meds, where each vertex $v \in V$ is time-independently mapped to a point in the plane and each edge $(u,v) \in E$ is mapped to two fragments of the segment $\overline{uv}$, one incident to $u$ and one incident to~$v$. We call these two subsegments \emph{stubs}, which change length smoothly over~time.

Depending on the length changing process, 
\meds are \emph{symmetric} if at any time the two stubs of any edge have the same length. Previous studies show that symmetry allows the user to efficiently match stubs belonging to the same edge~\cite{DBLP:conf/iisa/BinucciLMT16}. 
We can describe the stub length of an edge $(u,v)$ in a symmetric \med by a  \emph{stub length ratio} function $\delta_{(u,v)}(t)$ that describes the ratio between the stub length and the length of the segment $\overline{uv}$. If the minimum of the stub length ratio function of all edges is the same value $\delta_0$, we call the drawing \emph{homogeneous}. We consider only \emph{symmetric} and \emph{homogeneous} \meds, also known as $\delta_0$-\shmeds, as it is known that symmetric and homogeneous \peds are the easiest to read~\cite{DBLP:conf/iisa/BinucciLMT16}. 

Similar to  previous studies~\cite{DBLP:conf/gd/Misue22,DBLP:conf/gd/MisueA19}, we consider only stub length ratio functions that morph stub lengths between two distinct states (namely $\delta_0$ and $1/2$) such that the transition between both states is 
\emph{strictly monotone}. Namely, we allow such a smooth transition to happen several times for each edge, starting from specific time frames $T_{start}(u,v) \subset T$ during which $\delta_{(u,v)}(t)=\delta_0$. After some predefined elapsed time $\tau_{(u,v)}$ the edge is represented as a straight-line segment, i.e., $\delta_{(u,v)}(t)=1/2$. Afterwards, for some duration $\tau_{1/2}$, edge $(u,v)$ is completely visible ensuring that the user can see the full edge for some frames in a rendered animation as suggested by Misue~\cite{DBLP:conf/gd/Misue22}. Following this small delay, the animation reverts and $\delta_{(u,v)}=\delta_0$ again after time $\tau_{(u,v)}$. Due to the relation of both parts of the animation, in the following we only describe the first part of the animation. As in previous studies~\cite{DBLP:conf/gd/Misue22}, we use the same type of animation for all morphs of an edge, i.e., $\tau_{(u,v)}$ is assumed to be constant. To facilitate distinguishing between several runs of an animation, we allow a new animation to start only after $\tau_{distinct}$ time has elapsed. 

If we consider the animations of two edges $e_1$ and $e_2$, we may observe that an \emph{avoidable} crossing (between two stubs of length ratio strictly greater than $\delta_0$) can occur if their animations are not scheduled accordingly. We want to avoid that the stubs of $e_1$ and $e_2$ ever meet at their avoidable crossing as otherwise they can be perceived as the same object according to Gestalt principles~\cite{rusu2011using}. To counteract this from happening, we also allow $e_1$ to pass through the same point $e_2$ has passed through (and vice-versa) only after $\tau_{distinct}$ time has elapsed.

\paragraph{Animation Parameters.} 
We want to describe the morph from $\delta_{(u,v)}(t)=\delta_0$ to $\delta_{(u,v)}(t+\tau_{(u,v)})=1/2$  with two easier to understand parameters that we can fix for the \emph{entire} \med so that the aesthetic effect of all edge morphs is similar.

The first parameter is the \emph{speed} of the stub length morphing animation. In the previous studies on \med~\cite{DBLP:conf/gd/Misue22,DBLP:conf/gd/MisueA19}, the speed was a constant $\sigma$ holding for all edges. We can extend the speed parameter to morphings with
\emph{non-constant speed} by bounding the \emph{average morphing speed} $\sigma_a$ or the \emph{maximum morphing speed} $\sigma_m$. It is straight-forward to keep $\sigma_a$ fixed as for fixed $\sigma_a$, $\tau_{(u,v)}$ simply depends on the length of segment $\overline{uv}$. Bounding $\sigma_m$ is more tricky as it requires $\delta_{(u,v)}$ to be differentiable to be efficient to compute. Thus, in this paper, we only consider the {average morphing speed} $\sigma_a$ as a parameter.

In addition, we may want to consider animations with a non-constant speed. This may be desirable as humans are used to movements in nature that usually are at first gradually accelerating before reaching a maximum speed before slowing again before coming to a halt. This observation has been considered in multimedia design for a long time~\cite{penner} and gave rise to so-called \emph{easing} of animations~\cite{mdn,penner}. Namely, an \emph{easing function} $\eta: [0,1] \rightarrow [0,1]$ expresses for a given time elapsed ratio $\rho_t$  the progress of the animation $\rho_a$. In this regard, $\rho_t=0$ means that at the current time-frame the animation starts (in our case, $t = t_s$ for a $t_s \in T_{start}(u,v)$) whereas $\rho_t=1$ means that at the current time-frame the animation finishes (in our case, $t =t_s + \tau_{(u,v)}$). Meanwhile  $\rho_a=0$ means that the animation has just started (in our case, the stub length ratio is $\delta_0$) whereas $\rho_a=1$ means that it has been fully displayed (in our case, the stub length ratio has become $1/2$). For our purposes, we want that $\eta$ is \emph{strictly monotone} and \emph{invertible} \emph{on the interval $[0,1]$} to be able to apply the algorithm of Misue and Akasaka~\cite{DBLP:conf/gd/MisueA19}(refer to Appendix~\ref{app:misueAkasaka} for a description of the algorithm).

\section{User Study}\label{se:user-study}

\subsection{Experimental Conditions}

\begin{table}[t]
\begin{minipage}[c]{0.375\textwidth}
    \centering
    \includegraphics[scale=1]{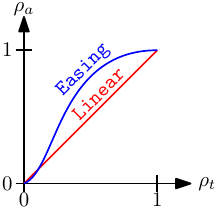}
    \captionof{figure}{Easing functions.}
    \label{fig:easing}
\end{minipage}
\hfill
\begin{minipage}[c]{0.6\textwidth}
\caption{Parameters used in our experiments.}
\centering
\begin{tabular}{
>{\columncolor[HTML]{EFEFEF}}c 
>{\columncolor[HTML]{CBCEFB}}c 
>{\columncolor[HTML]{FFCCC9}}c 
>{\columncolor[HTML]{FFFFC7}}c 
>{\columncolor[HTML]{CBFBCE}}c}
\hline
\cellcolor[HTML]{DAE8FC}\textbf{Model:}         & \cellcolor[HTML]{DAE8FC}\texttt{SlowLin}        & \cellcolor[HTML]{DAE8FC}\texttt{SlowEas}   & \cellcolor[HTML]{DAE8FC}\texttt{FastLin}        & \cellcolor[HTML]{DAE8FC}\texttt{FastEas}                                              \\ \hline
$\sigma_a$                                                              & 100 px/s   & 100 px/s                                         & 200 px/s     & 200 px/s                                                                                 \\
$\eta$                                                                            & \texttt{Linear} & \texttt{Easing} & \texttt{Linear} & \texttt{Easing} \\
$\tau_{1/2}$             &                   \multicolumn{4}{c}{ \cellcolor[HTML]{FFFFFF}  100 ms}                                                                                                               \\
$\tau_{distinct}$                                                        & \multicolumn{4}{c}{ \cellcolor[HTML]{FFFFFF}  50 ms}                                                                                           \\ $\delta_{0}$                                                        & \multicolumn{4}{c}{ \cellcolor[HTML]{FFFFFF}  1/4}                                                                                           \\\hline
\end{tabular}
\label{tab:models}
\end{minipage}
\end{table}

We investigate how morphing speed and easing function influence the readability of \meds. 
In the first study on \med, $\sigma_a$ was chosen at $10^\circ$/s to be as fast as possible\footnote{
    The metric $^\circ$/s refers to the temporal change of the visual angle, which is the size~of the image of the object on the observer's retina. Naturally, this metric requires a controlled environment which we believe to conflict with a variance in user preferences.
} while remaining human-tractable~\cite{DBLP:conf/gd/MisueA19} whereas in the follow-up study~\cite{DBLP:conf/gd/Misue22} it was significantly reduced to 100 px/s. We intend to evaluate \med as a generally applicable tool for graph visualizations in practical applications, where the concrete usage depends on user preferences. As a result, we believe that px/s is a more useful metric than $^\circ$/s as the latter metric requires to control the relative positioning of the user towards the screen\footnote{The concrete effect of a speed measured in px/s vastly depends on the screen resolutions. Hence, we fixed the speed assuming a resolution of 1920 px $\times$ 1080 px and scaled the drawings so to cover the same fraction of the screen size on any resolution.}. We investigate two speeds:
\begin{compactitem}
    \item[\texttt{Slow.}] Here, we set $\sigma_a=100$ px/s as in~\cite{DBLP:conf/gd/Misue22}.
    \item[\texttt{Fast.}] Here, we set $\sigma_a=200$ px/s so to have a significantly faster speed as comparison which allows to check how a shorter animation duration at the cost of higher speed influences readability.
\end{compactitem}

\noindent In previous studies on \med~\cite{DBLP:conf/gd/Misue22,DBLP:conf/gd/MisueA19}, as an easing function, the function $
\rho_a=\eta(\rho_t)=\rho_t$, which is also known as the \texttt{Linear} easing function, has been used. We believe that the aesthetic appeal of \meds can be improved by having a non-constant speed allowing for a more natural appearance of the movement of the stubs. As a side effect, edges can be displayed \emph{almost fully drawn} for a longer duration which may allow users to more efficiently observe them; see Fig.~\ref{fig:easing}. 
There is a plethora of different easing functions in the literature~\cite{penner}. Currently, functions described by \emph{cubic Bézier curves} appear as the web-design standard~\cite{mdn,DBLP:conf/cgi/IzdebskiKS20} with a particular focus on the function described by points $P_0=(0,0)$, $P_1=(0.25,0.1)$, $P_2=(0.25,1)$ and $P_3=(1,1)$ which is commonly known also as \texttt{Easing}; see Fig.~\ref{fig:easing}. It is worth noting that cubic Bézier curves are defined as a function that maps a parameter $p$ to the tuple $(\rho_t,\rho_a)$. As a result, it is not straight-forward to compute $\rho_a$ in terms of $\rho_t$~\cite{DBLP:conf/cgi/IzdebskiKS20} and implementations typically solve this problem numerically~\cite{asawicki}. For our purposes, we oriented ourselves at the implementation used in the \textsc{Firefox} web browser~\cite{firefox}.

The combination of speed (\texttt{Slow}/\texttt{Fast}) and easing function (\texttt{Linear}/\texttt{Easing}) gives rise to the four models, i.e., the experimental conditions, that we compare in our study; see Table~\ref{tab:models}. Note that in all models we set $\tau_{1/2}=100$ ms (as in~\cite{DBLP:conf/gd/Misue22}) and $\tau_{distinct}=50$ ms. Further, we used $\delta_0=1/4$ as in~\cite{DBLP:conf/gd/MisueA19}.

\subsection{Tasks and Hypotheses}

\begin{figure}[t!]
\centering
    \includegraphics[width=.45\textwidth]{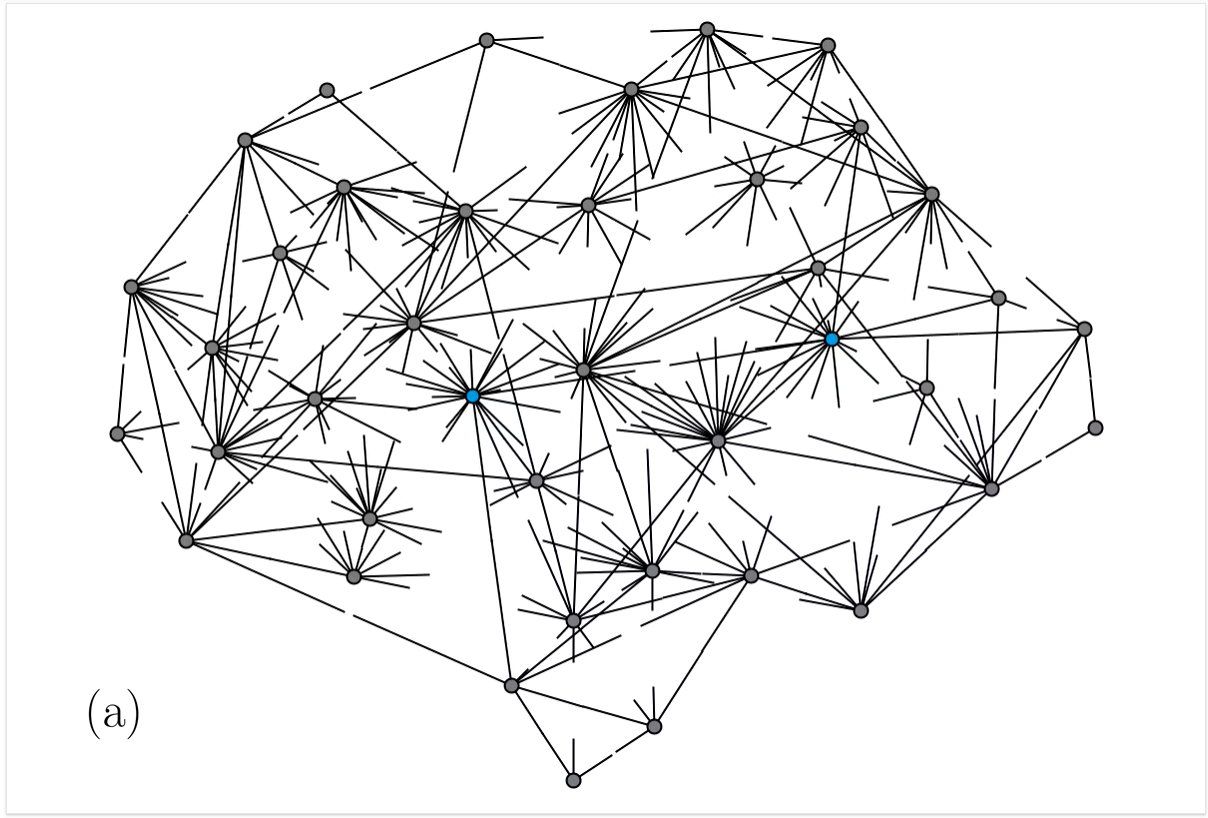}\label{fi:trials-a}
    \hfil
    \includegraphics[width=.45\textwidth]{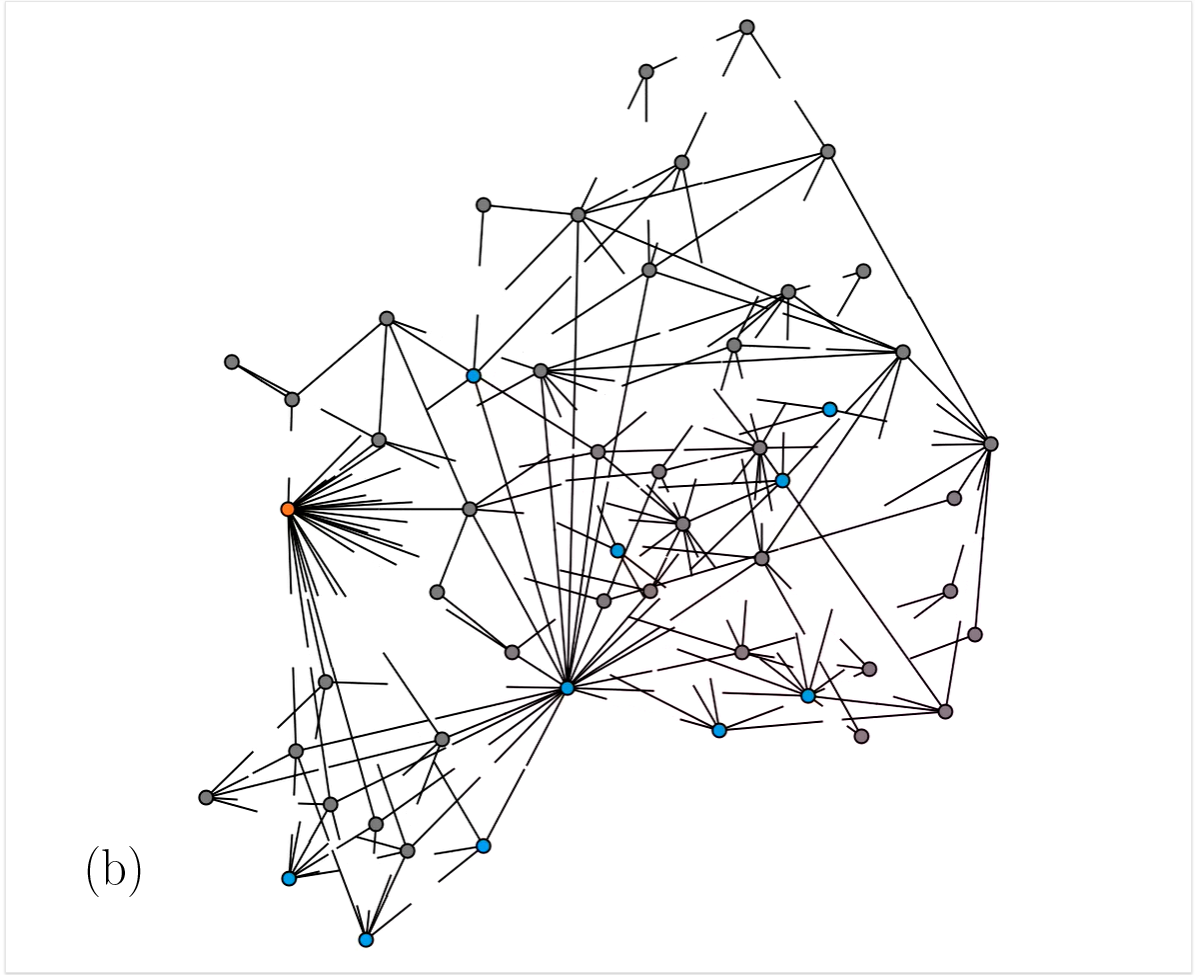}\label{fi:trials-b}
    \hfil
    \includegraphics[width=.45\textwidth]{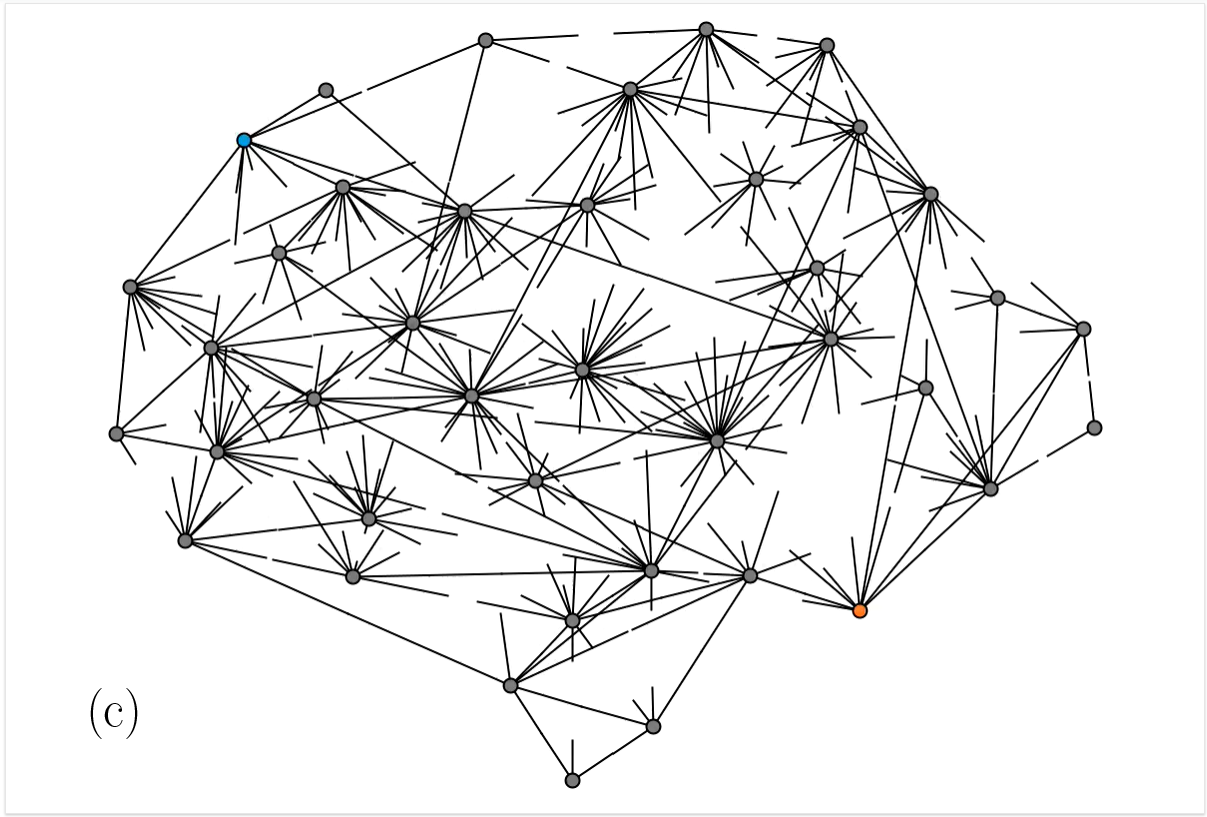}\label{fi:trials-c}
    \hfil
    \includegraphics[width=.45\textwidth]{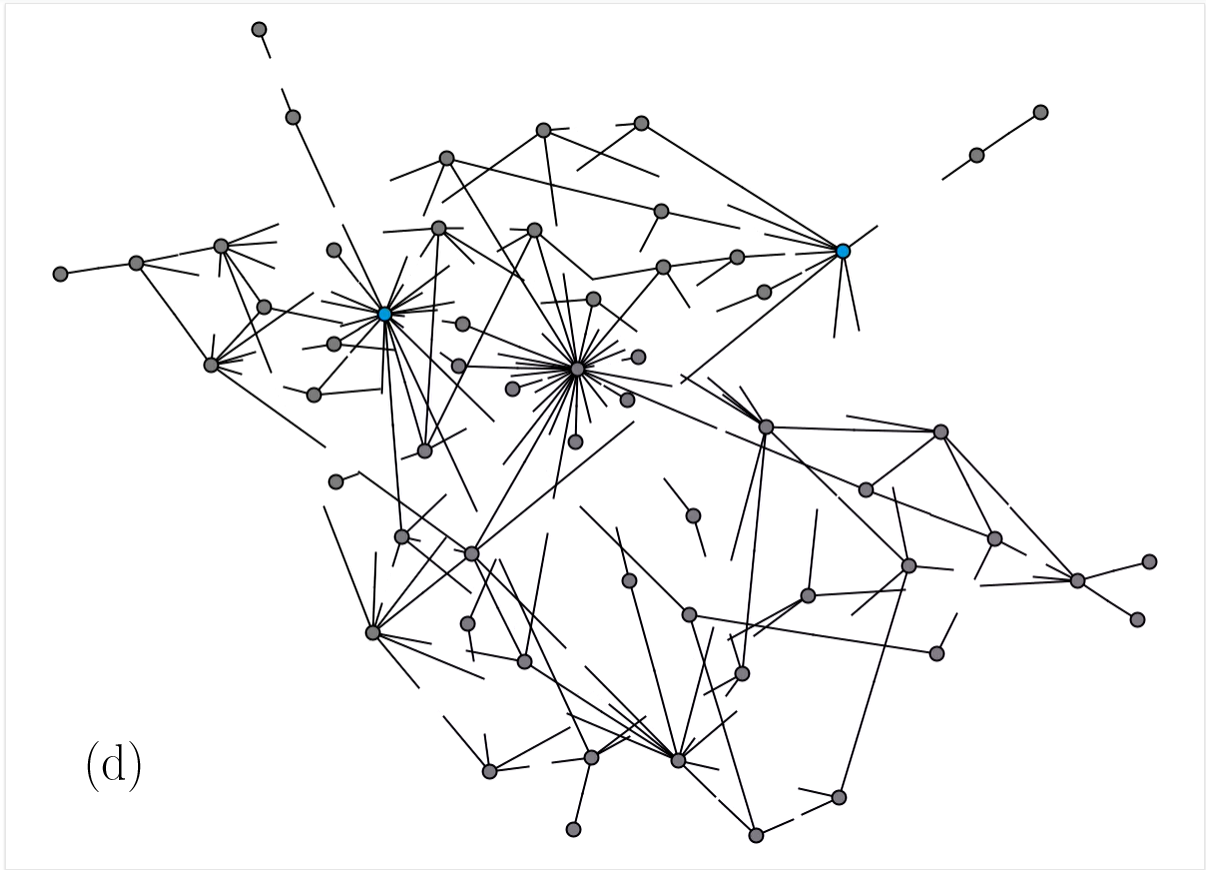}\label{fi:trials-d}
    
   \includegraphics[width=.45\textwidth]{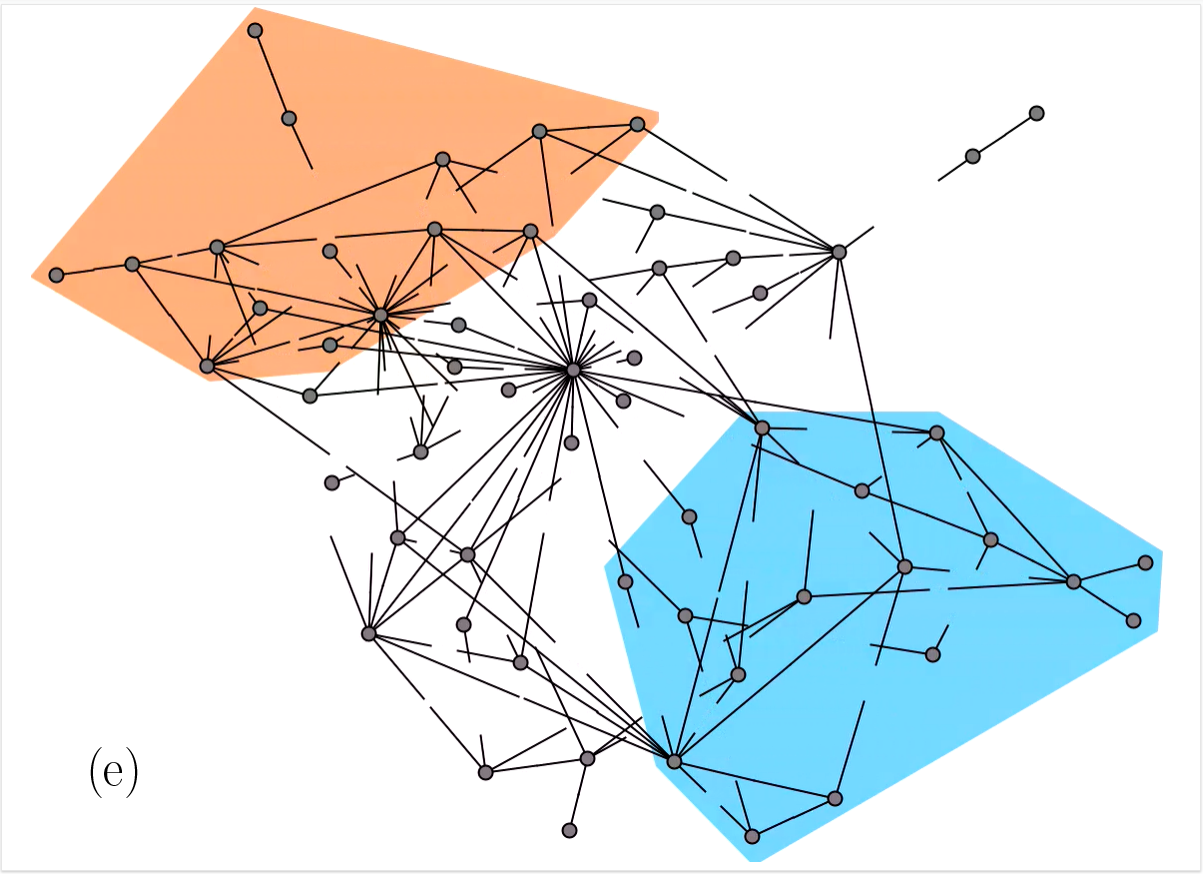}\label{fi:trials-e}
    \caption{Examples of trials: (a)~\textsf{T1~(Adjacency)}, (b)~\textsf{T2~(NeighborhoodSize)}, (c)~\textsf{T3~(PathLength)} with $k=2$, (d)~\textsf{T4~(CommonNeighbors)}, (e)~\textsf{T5~(InterRegionEdges)}.}
    \label{fi:trials}
\end{figure}

We defined five tasks, reported in Table~\ref{ta:tasks}, for which we also provide the classification according to the taxonomy by Lee et al.~\cite{lpsfh-ttgv-beliv2006}. Fig.~\ref{fi:trials} shows some examples of trials.
We designed the tasks so that they require to explore the drawing locally and globally, are easy to explain, can be executed in a reasonably short time, and can be easily measured. Also, most of them have already been used in previous graph visualization user studies (e.g.,~\cite{DBLP:journals/cgf/BinucciDKLM22,10004748,ojk-nlamoqni-tvcg2019,DBLP:journals/vlc/Purchase98}).
The main purpose~of our study is to evaluate the differences between our experimental conditions in terms of readability, understandability, and effectiveness. For this reason we defined \emph{interpretation tasks}, according to the top-level classification by Burch~et~al.~\cite{bhwpwh-saeegv-ieeeaccess2021}. 

\begin{table}[h]
\caption{Tasks used in our experiment.}\label{ta:tasks}
\centering
\begin{tabular}{|c|c|l|ll}
\cline{1-3}
\cellcolor[HTML]{DAE8FC}\textbf{Task} & \cellcolor[HTML]{DAE8FC}\textbf{Classification}                              & \cellcolor[HTML]{DAE8FC}\textbf{Description}                                                                                       &  &  \\ \cline{1-3}
\begin{tabular}[c]{@{}c@{}}\textsf{T1}\\\textsf{(Adjacency)}\end{tabular}                                             & \begin{tabular}[c]{@{}c@{}}topology-based\\ (adjacency)\end{tabular}         & \begin{tabular}[c]{@{}l@{}}Is there an edge that connects \\ the two \textbf{\myblue{blue}} nodes?\end{tabular}                                      &  &  \\ \cline{1-3}
\begin{tabular}[c]{@{}c@{}}\textsf{T2}\\\textsf{(NeighborhoodSize)}\end{tabular}                         & \begin{tabular}[c]{@{}c@{}}topology-based\\ (adjacency)\end{tabular}         & \begin{tabular}[c]{@{}l@{}}How many \textbf{\myblue{blue}} nodes are connected \\ with the \textbf{\myorange{orange}} node?\end{tabular}                                 &  &  \\ \cline{1-3}
\begin{tabular}[c]{@{}c@{}}\textsf{T3}\\\textsf{(PathLength)}\end{tabular}                               & \begin{tabular}[c]{@{}c@{}}topology-based\\ (connectivity)\end{tabular}      & \begin{tabular}[c]{@{}l@{}}Is there a path of length at most $k$ \\ that connects the \textbf{\myblue{blue}} node with the   \\\textbf{\myorange{orange}} node?\end{tabular} &  &  \\ \cline{1-3}

\begin{tabular}[c]{@{}c@{}}\textsf{T4}\\\textsf{(CommonNeighbors)}\end{tabular}                                  & \begin{tabular}[c]{@{}c@{}}topology-based\\ (common connection)\end{tabular} & \begin{tabular}[c]{@{}l@{}}How many nodes are connected \\ with both the \textbf{\myblue{blue}} nodes?\end{tabular}                                  &  &  \\ \cline{1-3}
\begin{tabular}[c]{@{}c@{}}\textsf{T5}\\\textsf{(InterRegionEdges)}\end{tabular}                     & overview                                                                     & \begin{tabular}[c]{@{}l@{}}How many edges directly connect \\ the two highlighted parts?\end{tabular}                              &  &  \\ \cline{1-3}
\end{tabular}
\end{table}

\smallskip
We formulated three experimental hypotheses about the effectiveness of the different conditions to support users
in the execution of analysis tasks: 
\begin{compactenum}
    \item[\textsf{H1.}] \texttt{Fast} speed and cubic Bézier curve \texttt{Easing} give lower response time than \texttt{Slow} speed and \texttt{Linear} easing.
    \item[\textsf{H2.}] \texttt{Slow} speed and cubic Bézier curve \texttt{Easing} give lower error rate than \texttt{Fast} speed and  \texttt{Linear} easing.
    \item[\textsf{H3.}] \texttt{Slow} speed and cubic Bézier curve \texttt{Easing} are preferred by the users rather than \texttt{Fast} speed and \texttt{Linear} easing.
\end{compactenum}

\noindent Our rationale behind \textsf{H1} is that the time used for morphing all edges with \texttt{Fast} speed is shorter than with \texttt{Slow} speed, and cubic Bézier curve \texttt{Easing} makes stubs touch for a longer time than with \texttt{Linear} easing. About \textsf{H2}, we think that the user can benefit from cubic Bézier curve \texttt{Easing} and \texttt{Slow} speed in terms of accuracy, since the movement of the stubs is more gradual and connections are visualized for a longer time than with \texttt{Linear} easing and \texttt{Fast} speed. 
Finally, about \textsf{H3} we think that users may find a \texttt{Fast} animation more irritating and distracting than a \texttt{Slow} one, and the movement of stubs with \texttt{Linear} easing less smooth than with cubic Bézier curve \texttt{Easing}.

\subsection{Stimuli}

We investigated our hypotheses on visualizations of real-world networks, more specifically drawn from the annual \textsc{GD Contest}\footnote{See \url{https://mozart.diei.unipg.it/gdcontest/} for further information.}:

    \texttt{boardgames.} This graph from the GD Contest 2023 contains the 40 most popular boardgames as nodes which are connected if fans of one game also liked the other one (data according to {\small \url{https://boardgamegeek.com/}}). 
    
    \texttt{marvel.} This bipartite graph from the GD Contest 2019  represents which characters from the Marvel Cinematic Universe are featured in which film. 
    
    \texttt{kpop.} This graph consists of a subset of the data from the GD Contest 2020 and shows the nodes with betweenness centrality at least 0.3 from the largest connected component of the subgraph induced by K-Pop bands and labels. Connections in this network showcase cooperations between the entities.

We chose the specific networks as they all have a moderate size reasonably displayable on a standard screen (40 to 58 nodes) and offer various densities (ranging from 2.12 to 5.35). We embedded the stimuli graphs using a force directed
algorithm available in the \texttt{D3.js} library~\cite{d3}. The resulting drawings roughly required the same area; see Table~\ref{tab:stimuli}.
%

In our drawings, nodes are represented as disks of radius 7 px filled gray, blue or orange (where the colored variants are used to highlight nodes of interest for the tasks), while the boundaries of the disks and edges are drawn black and 2 px wide. The animations were scheduled using the algorithm of Misue and Akasaka~\cite{DBLP:conf/gd/MisueA19} with an improvement of~\cite{DBLP:conf/gd/Misue22} (see Appendix~\ref{app:misueAkasaka} for details) and rendered at 30 fps. The resulting animations for \texttt{SlowEas} (\texttt{FastEas}) were on average $20\%$ ($11\%$, resp.) slower than the corresponding animations for \texttt{SlowLin} (\texttt{FastLin}, resp.). Comparing the different speeds, the animations for \texttt{SlowLin} (\texttt{SlowEas}) were on average $74\%$ ($87\%$, resp.) slower than the ones for \texttt{FastLin} (\texttt{FastEas}, resp.); see Table~\ref{tab:stimuli}. 
We produced a trial for each combination of network and task, so to obtain a total of $15$ trials for each experimental condition.
\begin{table}[t]

\caption{Data about the stimuli used in our experiment.}
\centering
\begin{tabular}{ccccccccc}
\hline
\rowcolor[HTML]{DAE8FC} 
\textbf{Stimulus}                                            & \textbf{Nodes} & \textbf{Edges} & \textbf{Density} & \textbf{Resolution} & \texttt{SlowLin} & \texttt{SlowEas} & \texttt{FastLin} & \texttt{FastEas} \\ \hline
\cellcolor[HTML]{EFEFEF}\texttt{boardgames} & 40                & 214      &     5.35     & 1030$\times$820 px & 7.79s &	8.74s	&4.25s &	4.87s
          \\
\cellcolor[HTML]{EFEFEF}\texttt{marvel}     & 52                & 152      &    2.92         & 846$\times$974 px &4.92s &	5.61s&	2.66s	&3.08s           \\ 
\cellcolor[HTML]{EFEFEF}\texttt{kpop}       & 58                & 123   &     2.12           & 1142$\times$858 px &5.31s &	7.09s	&3.43s	&3.55s           \\
\hline
\end{tabular}
\label{tab:stimuli}
\end{table}

\subsection{Experimental Process}

For our experiment, we opted for a between-subject design where each participant is exposed to one of the four conditions and hence to $15$ trials. Indeed, a within-subject design would imply that each user sees the same experimental object $20$ times, which makes it difficult to avoid the learning
\textcolor{blue}{
and the fatigue
}
effect. In addition, the between-subject study design allowed us to query subjective feedback from the users about a specific condition after a series of trials. The users performed an online test prepared with the LimeSurvey tool ({\small \url{https://www.limesurvey.org/}}). 

\noindent The phases of the survey are the following:
\begin{compactenum}
    \item we collect some information about the user;
    \item we assign the condition to the user by adopting a round robin approach;
    \item we present a video tutorial and some training questions;
    \item we present the $15$ trials in random order;
    \item we ask for some subjective feedback: $(i)$ Two 5-point Likert scale questions about the \textsf{beauty} of the drawings and the \textsf{easiness} of the questions, $(ii)$ one question about the perceived \textsf{speed} of the movements,
    $(iii)$ a question asking users if they found the tasks \textsf{tiring}, and $(iv)$ an optional free-form feedback. 
\end{compactenum}

\noindent
We used the \textsf{gdnet}, \textsf{ieee\_vis}, and \textsf{infovis} mailing lists to recruit participants. Also, we involved our engineering and CS students of the universities of Perugia and T\"ubingen.

\section{Results and Discussion}\label{se:results}

We collected questionnaires from 84 participants
(21 per condition);
see Table~\ref{ta:participants} for some details.

\begin{table}[tb]
\caption{Data about the participants.}
\centering
\begin{tabular}{|llll|
>{\columncolor[HTML]{EFEFEF}}l l|ll}
\hline
\multicolumn{4}{|c|}{\cellcolor[HTML]{DAE8FC}\textbf{Gender}}                                                                                                                                                                                                                                                                                 & \multicolumn{2}{c|}{\cellcolor[HTML]{DAE8FC}\textbf{Age}}          & \multicolumn{2}{c|}{\cellcolor[HTML]{DAE8FC}\textbf{Screen size}}                                   \\ \hline
\multicolumn{1}{|c|}{\cellcolor[HTML]{EFEFEF}{\color[HTML]{000000} \textit{Woman}}} & \multicolumn{1}{c|}{\cellcolor[HTML]{EFEFEF}{\color[HTML]{000000} \textit{Man}}} & \multicolumn{1}{c|}{\cellcolor[HTML]{EFEFEF}{\color[HTML]{000000} \textit{Transgender}}} & \multicolumn{1}{c|}{\cellcolor[HTML]{EFEFEF}{\color[HTML]{000000} \textit{Prefer not to say}}} & \multicolumn{1}{c|}{\cellcolor[HTML]{EFEFEF}\textit{18-24}} & 21\% & \multicolumn{1}{c|}{\cellcolor[HTML]{EFEFEF}\textit{\textless{}13"}}    & \multicolumn{1}{c|}{2\%}  \\ \hline
\multicolumn{1}{|c|}{\cellcolor[HTML]{FFFFFF}{\color[HTML]{000000} 31\%}}           & \multicolumn{1}{c|}{\cellcolor[HTML]{FFFFFF}{\color[HTML]{000000} 63\%}}         & \multicolumn{1}{c|}{\cellcolor[HTML]{FFFFFF}{\color[HTML]{000000} 1\%}}                  & \multicolumn{1}{c|}{\cellcolor[HTML]{FFFFFF}{\color[HTML]{000000} 5\%}}                        & \multicolumn{1}{c|}{\cellcolor[HTML]{EFEFEF}\textit{25-29}} & 26\% & \multicolumn{1}{c|}{\cellcolor[HTML]{EFEFEF}\textit{13"}}               & \multicolumn{1}{c|}{13\%} \\ 
\hline
\multicolumn{4}{|c|}{\cellcolor[HTML]{DAE8FC}\textbf{Educational level}}                                                                                                                                                                                                                                                                      & \multicolumn{1}{c|}{\cellcolor[HTML]{EFEFEF}\textit{30-34}} & 21\% & \multicolumn{1}{c|}{\cellcolor[HTML]{EFEFEF}\textit{14"}}               & \multicolumn{1}{c|}{21\%} \\ \hline
\multicolumn{1}{|c|}{\cellcolor[HTML]{EFEFEF}\textit{High school}}           & \multicolumn{1}{c|}{\cellcolor[HTML]{EFEFEF}\textit{Bachelor}}          & \multicolumn{1}{c|}{\cellcolor[HTML]{EFEFEF}\textit{Master}}                    & \multicolumn{1}{c|}{\cellcolor[HTML]{EFEFEF}\textit{Doctoral degree}}                          & \multicolumn{1}{c|}{\cellcolor[HTML]{EFEFEF}\textit{35-39}} & \multicolumn{1}{c|}{8\%}  & \multicolumn{1}{c|}{\cellcolor[HTML]{EFEFEF}\textit{15"}}               & \multicolumn{1}{c|}{20\%} \\ 
\hline
\multicolumn{1}{|c|}{14\%}                                                          & \multicolumn{1}{c|}{27\%}                                                        & \multicolumn{1}{c|}{30\%}                                                                & \multicolumn{1}{c|}{29\%}                                                                      & \multicolumn{1}{c|}{\cellcolor[HTML]{EFEFEF}\textit{40-44}} & \multicolumn{1}{c|}{7\%}  & \multicolumn{1}{c|}{\cellcolor[HTML]{EFEFEF}\textit{\textgreater{}15"}} & \multicolumn{1}{c|}{38\%} \\ \hline
\multicolumn{4}{|c|}{\cellcolor[HTML]{DAE8FC}\textbf{Expertise}}                                                                                                                                                                                                                                                                              & \multicolumn{1}{l|}{\cellcolor[HTML]{EFEFEF}\textit{45-49}} & \multicolumn{1}{c|}{5\%}  & \multicolumn{1}{l|}{\cellcolor[HTML]{EFEFEF}\textit{No answer}}         & \multicolumn{1}{c|}{5\%}  \\ \hline
\multicolumn{1}{|c|}{\cellcolor[HTML]{EFEFEF}\textit{None}}                         & \multicolumn{1}{c|}{\cellcolor[HTML]{EFEFEF}\textit{Low}}                        & \multicolumn{1}{c|}{\cellcolor[HTML]{EFEFEF}\textit{Medium}}                             & \multicolumn{1}{c|}{\cellcolor[HTML]{EFEFEF}\textit{High}}                                     & \multicolumn{1}{c|}{\cellcolor[HTML]{EFEFEF}\textit{50-59}} & \multicolumn{1}{c|}{5\%}  &                                                                         &                           \\ 
\hhline{------}
\multicolumn{1}{|c|}{12\%}                                                          & \multicolumn{1}{c|}{31\%}                                                        & \multicolumn{1}{c|}{26\%}                                                                & \multicolumn{1}{c|}{31\% }                                                                     & \multicolumn{1}{c|}{\cellcolor[HTML]{EFEFEF}\textit{60+}}   & \multicolumn{1}{c|}{6\%}  &                                                                         &                           \\ 
\hhline{------}
\end{tabular}
\label{ta:participants}
\end{table}

\subsection{Quantitative Results}
For each question, we recorded answers and response times of all participants. 
For \textsf{T1~(Adjacency)} and \textsf{T3~(PathLength)}, the error rate of a user is computed as the ratio between the number of wrong answers and the total number of questions, that is $3$ for each task.  For \textsf{T2~(NeighborhoodSize)}, \textsf{T4~(CommonNeighbors)}, and \textsf{T5~(InterRegionEdges)}, the error on a question is computed as $1-\frac{1}{1+|v_u-v_c|}$, where $v_u$ is the value given by the user and $v_c$ is the correct value.

\begin{figure}[tb]
\centering
    \includegraphics[width=.49\textwidth]{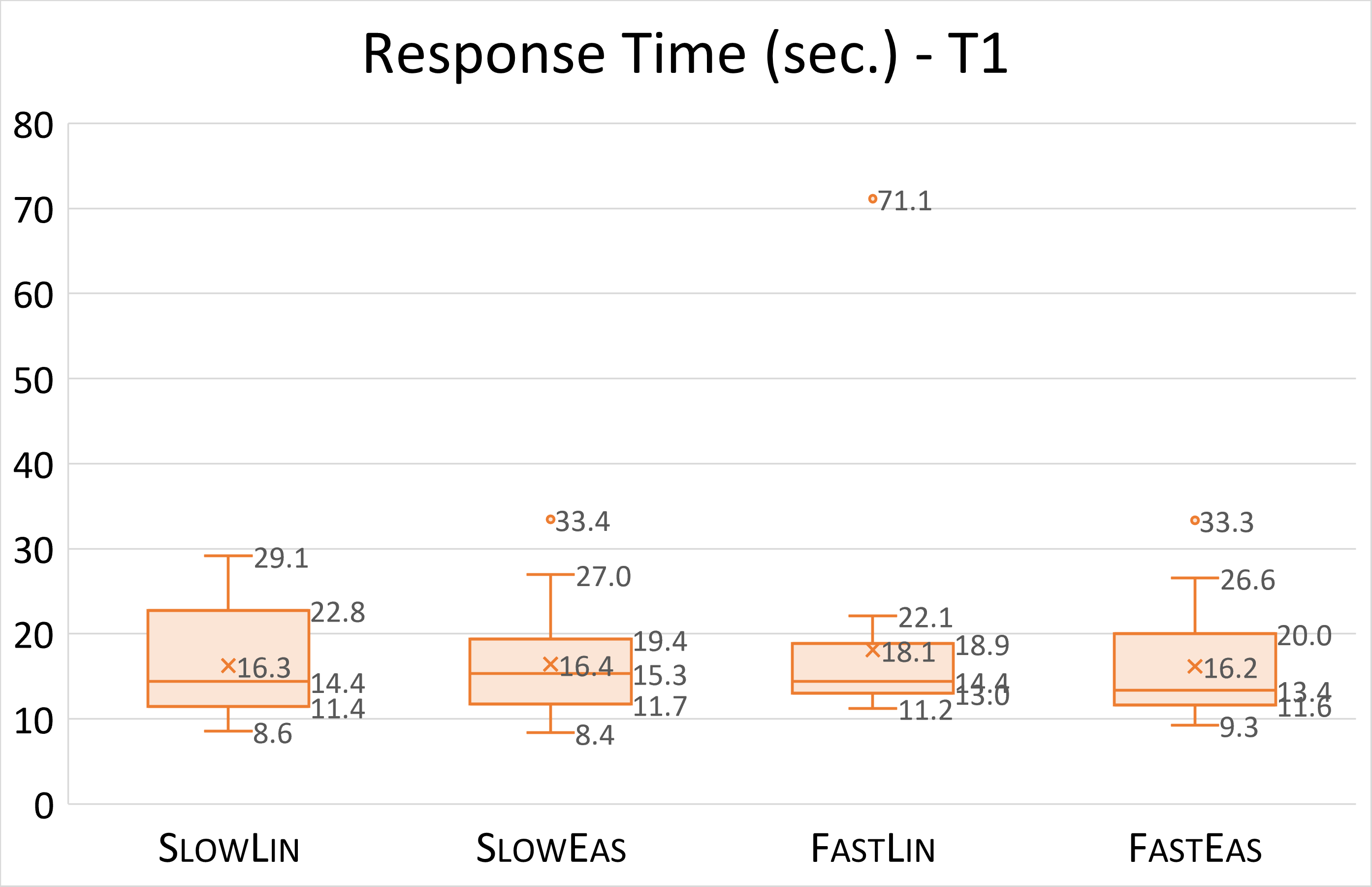}\label{fi:rt-a}
    \hfil
    \includegraphics[width=.49\textwidth]{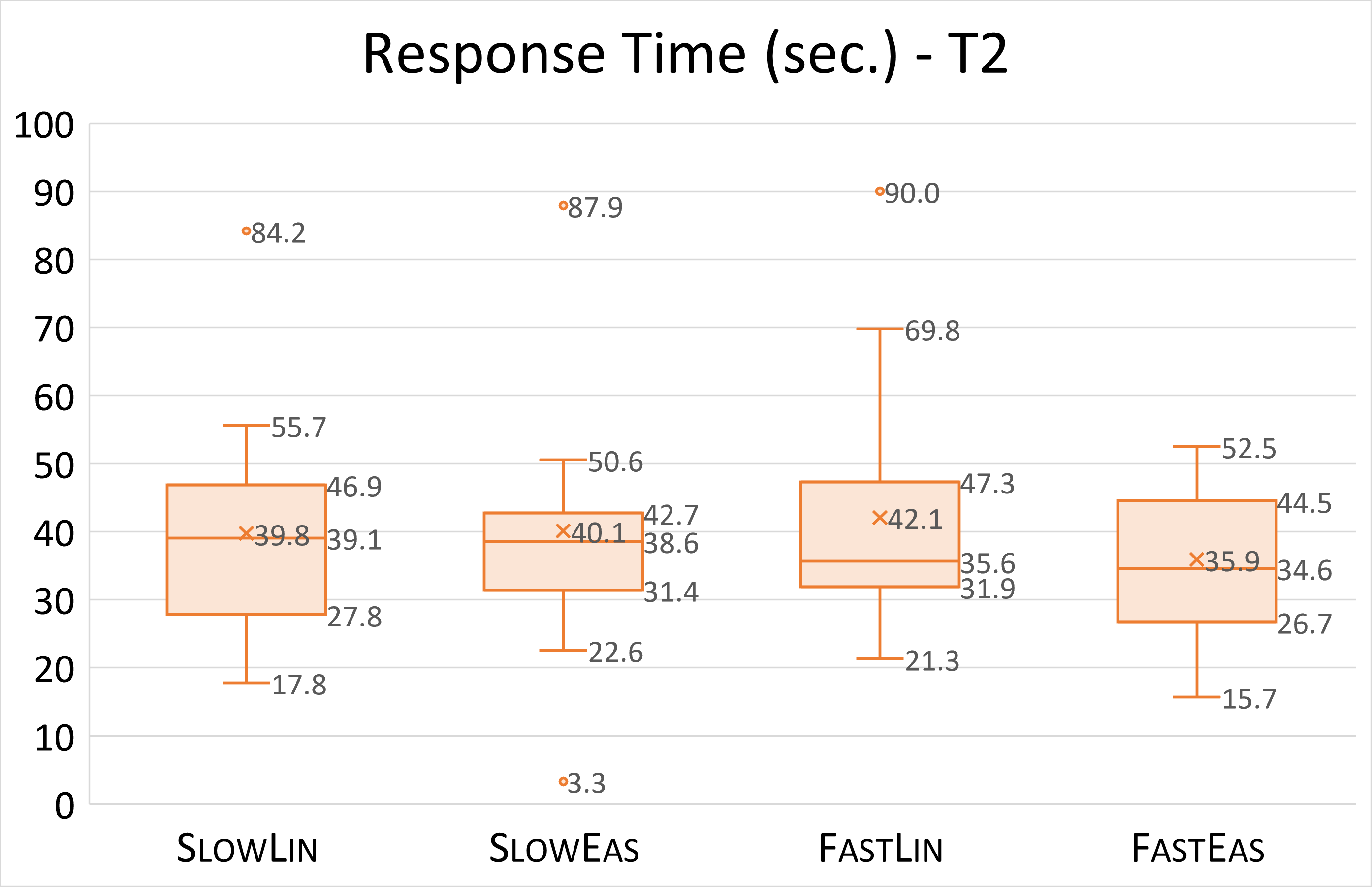}\label{fi:rt-b}
    \hfil
    \includegraphics[width=.49\textwidth]{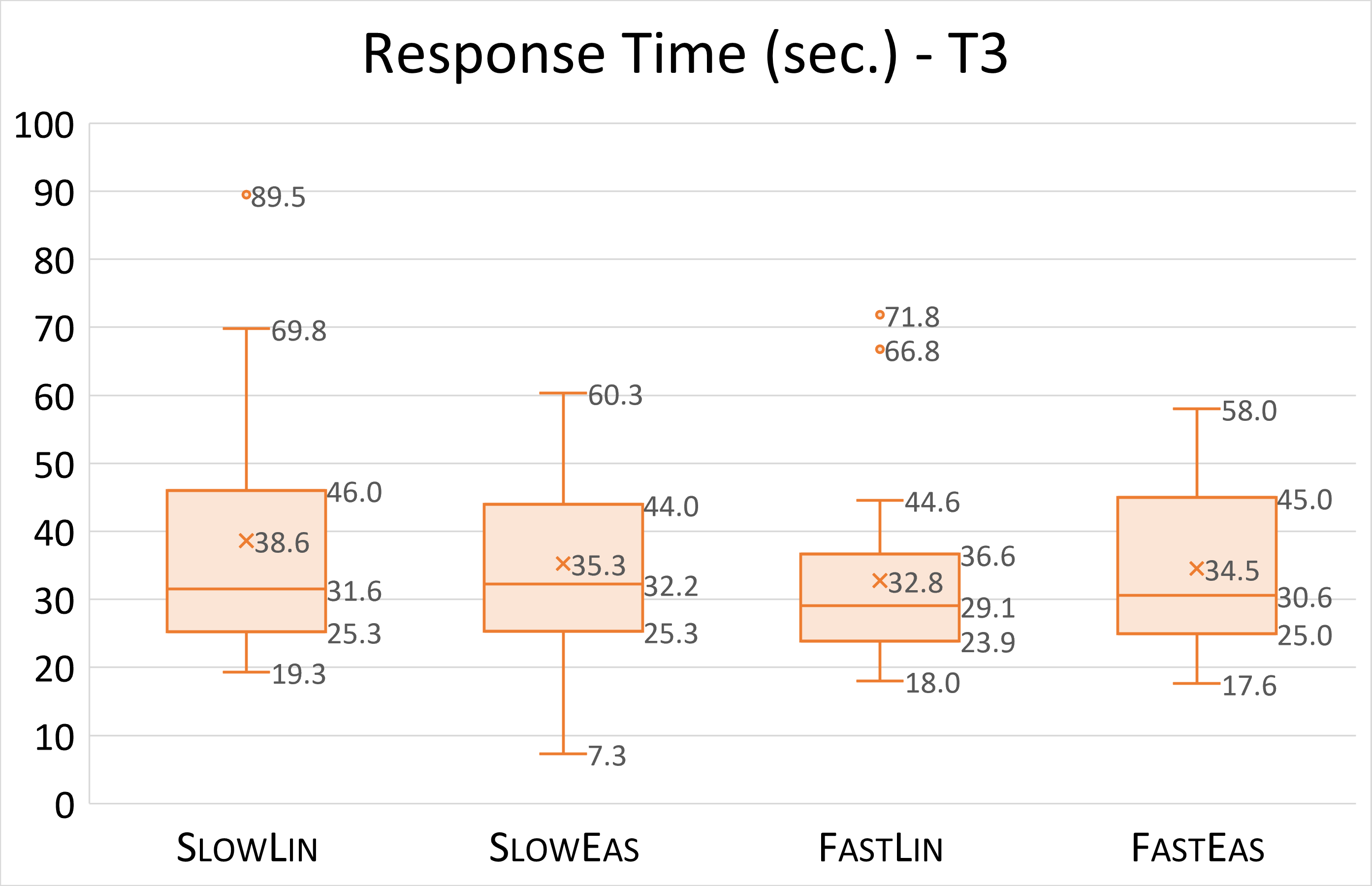}\label{fi:rt-c}
    \hfil
    \includegraphics[width=.49\textwidth]{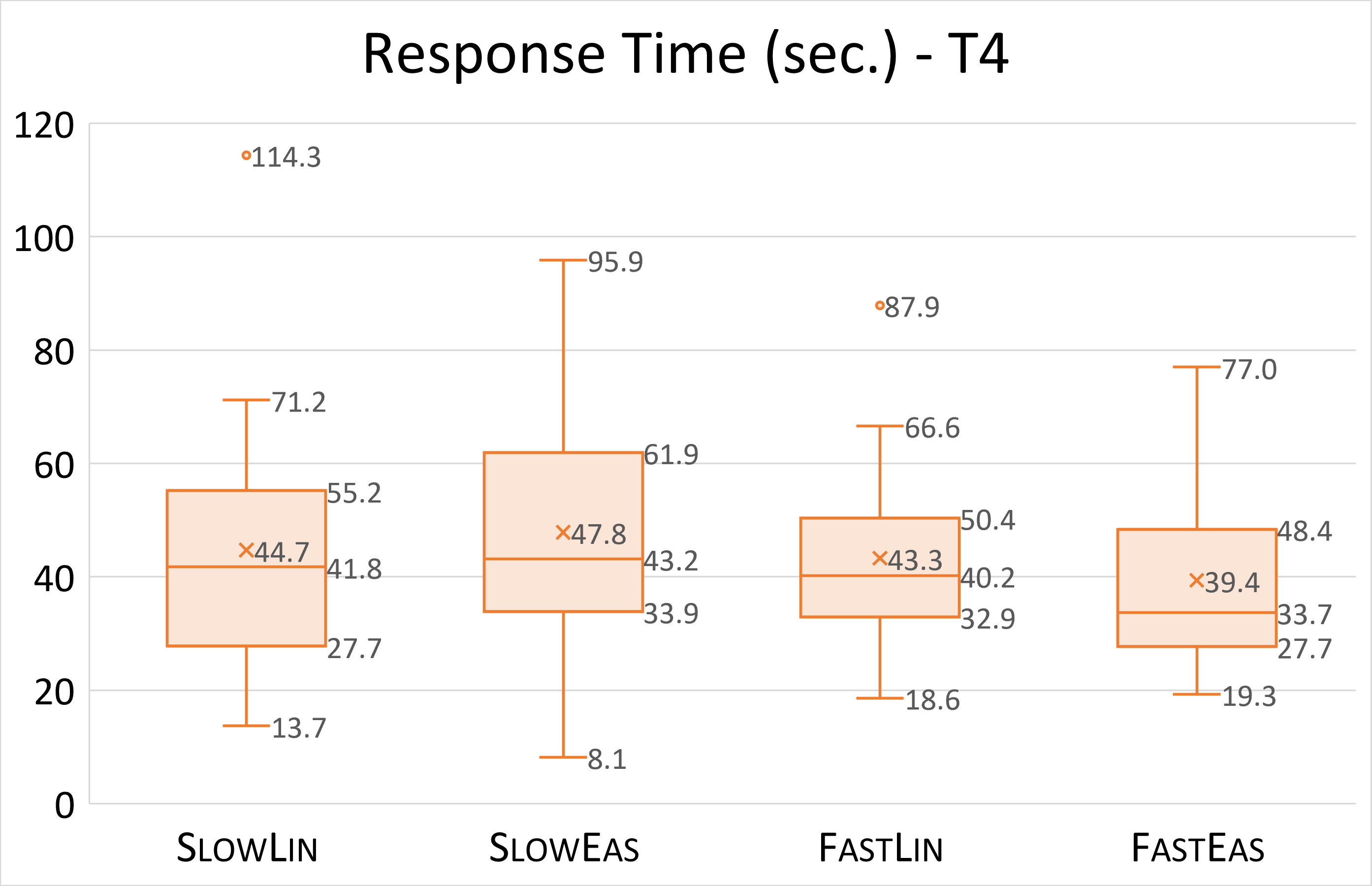}\label{fi:rt-d}
    \hfil
    \includegraphics[width=.49\textwidth]{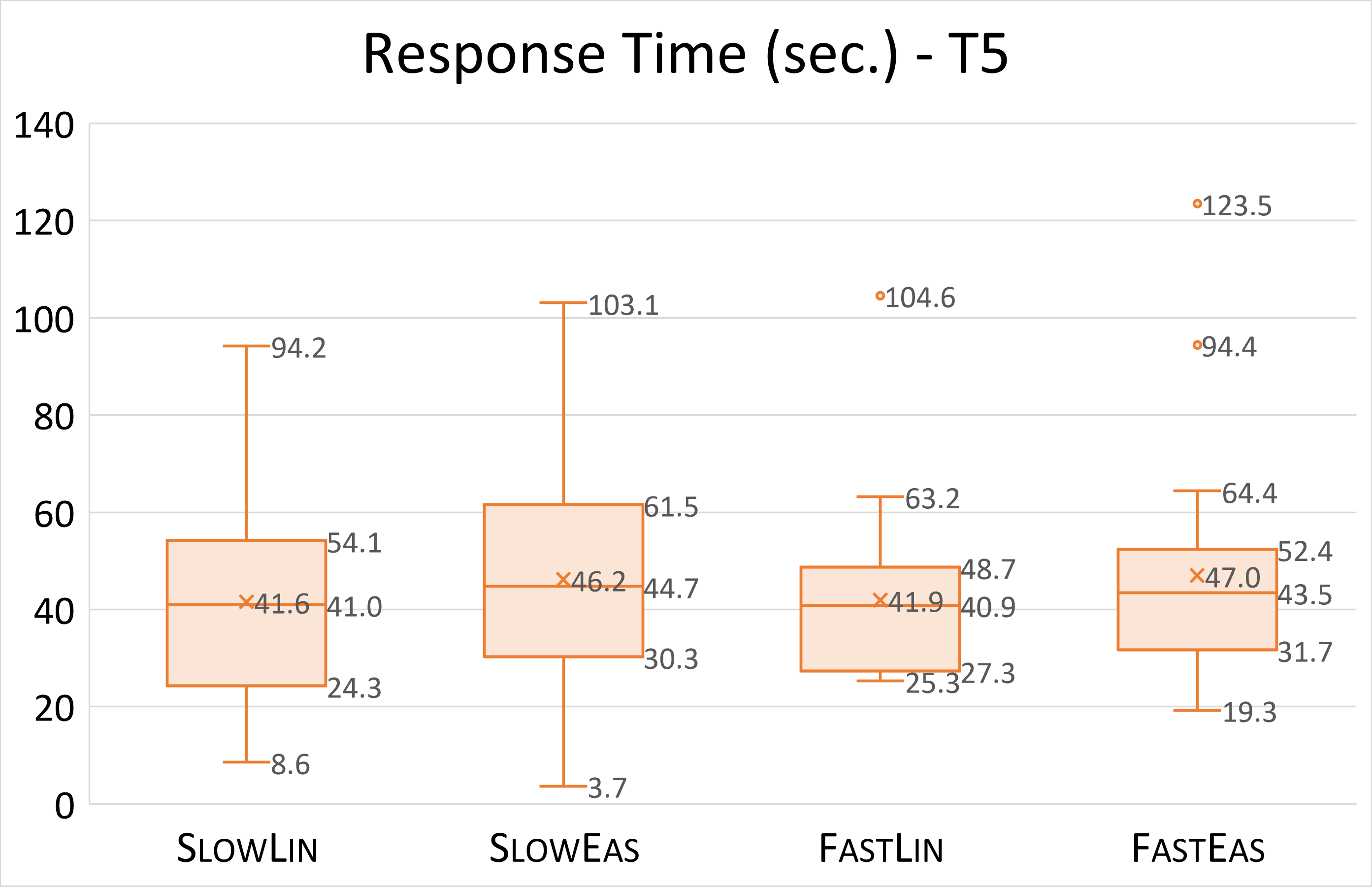}\label{fi:rt-e}
    \hfil
    \includegraphics[width=.49\textwidth]{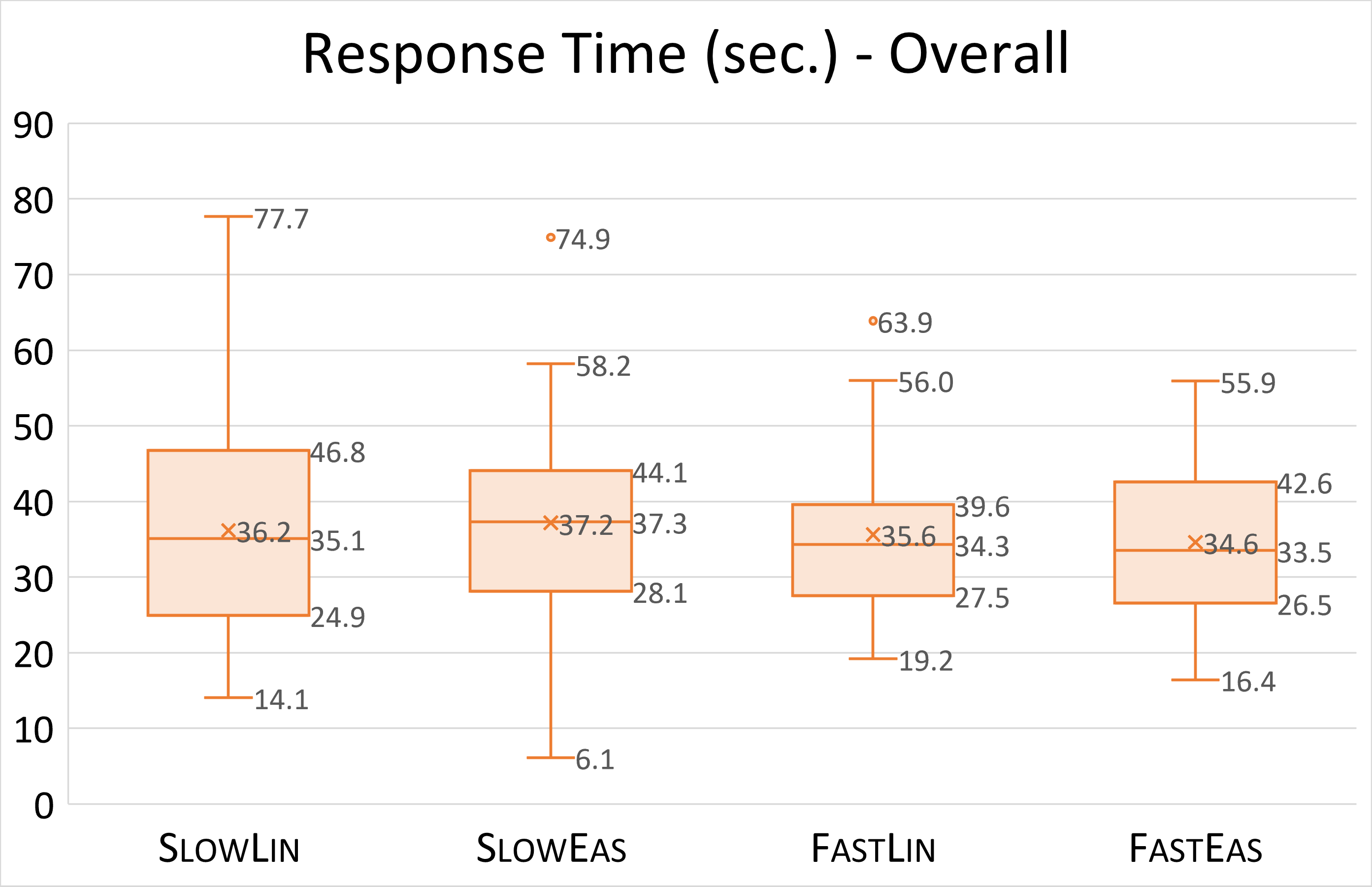}\label{fi:rt-f}
    \caption{Response time aggregated by task.}
    \label{fi:rt-box}
\end{figure}

\begin{figure}[tb]
\centering
    \includegraphics[width=.49\textwidth]{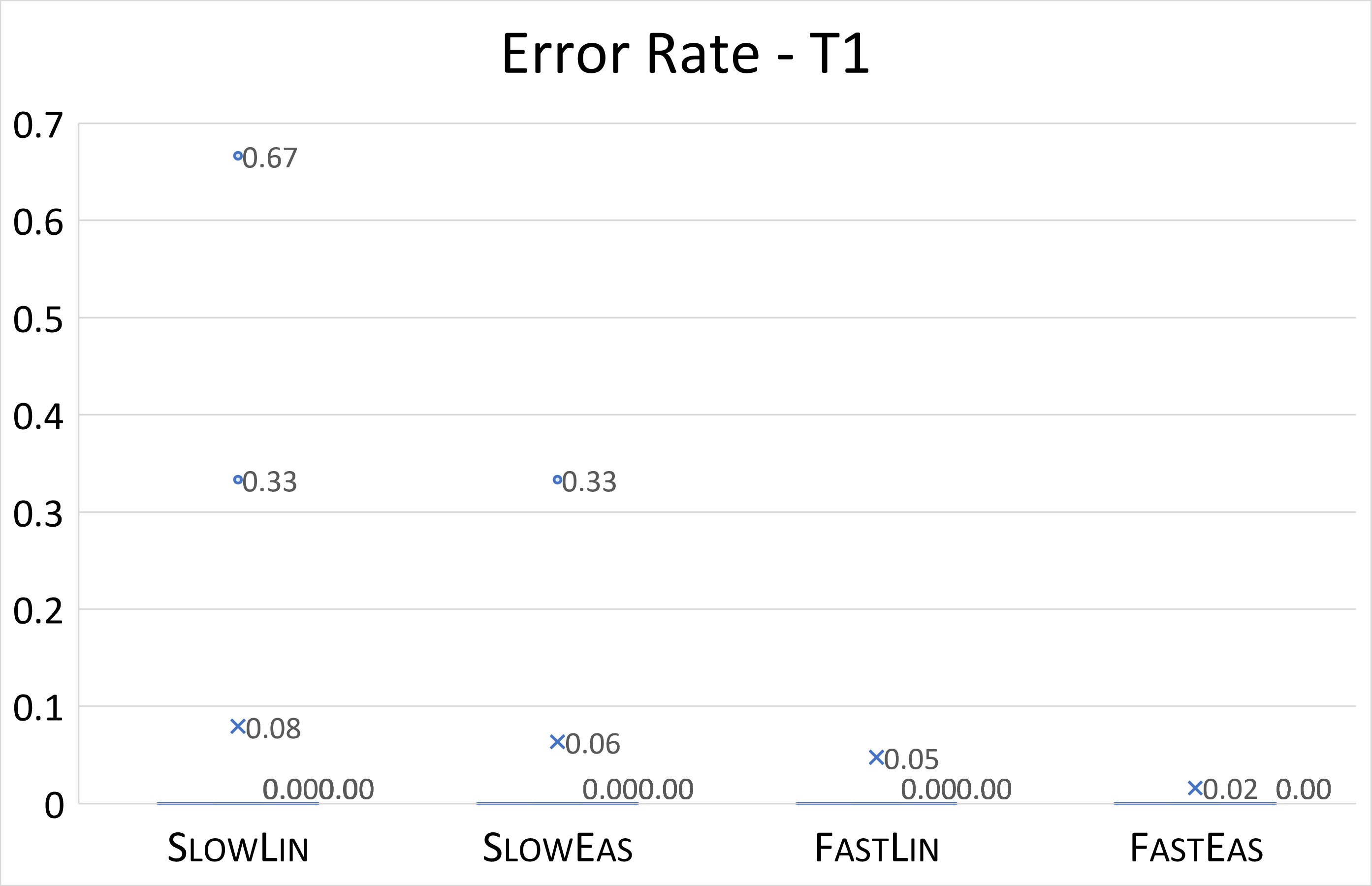}\label{fi:er-a}
    \hfil
    \includegraphics[width=.49\textwidth]{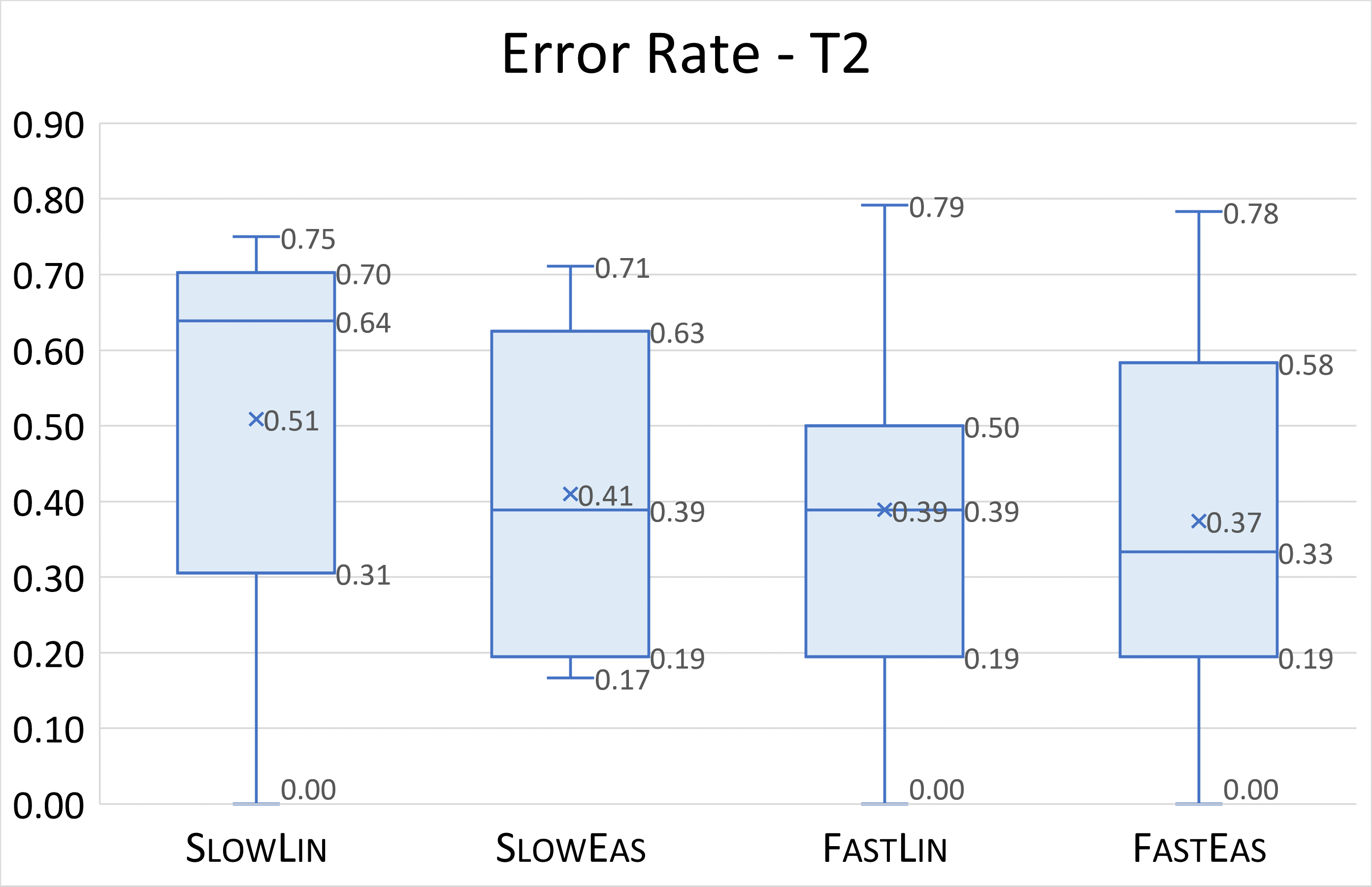}\label{fi:er-b}
    \hfil
    \includegraphics[width=.49\textwidth]{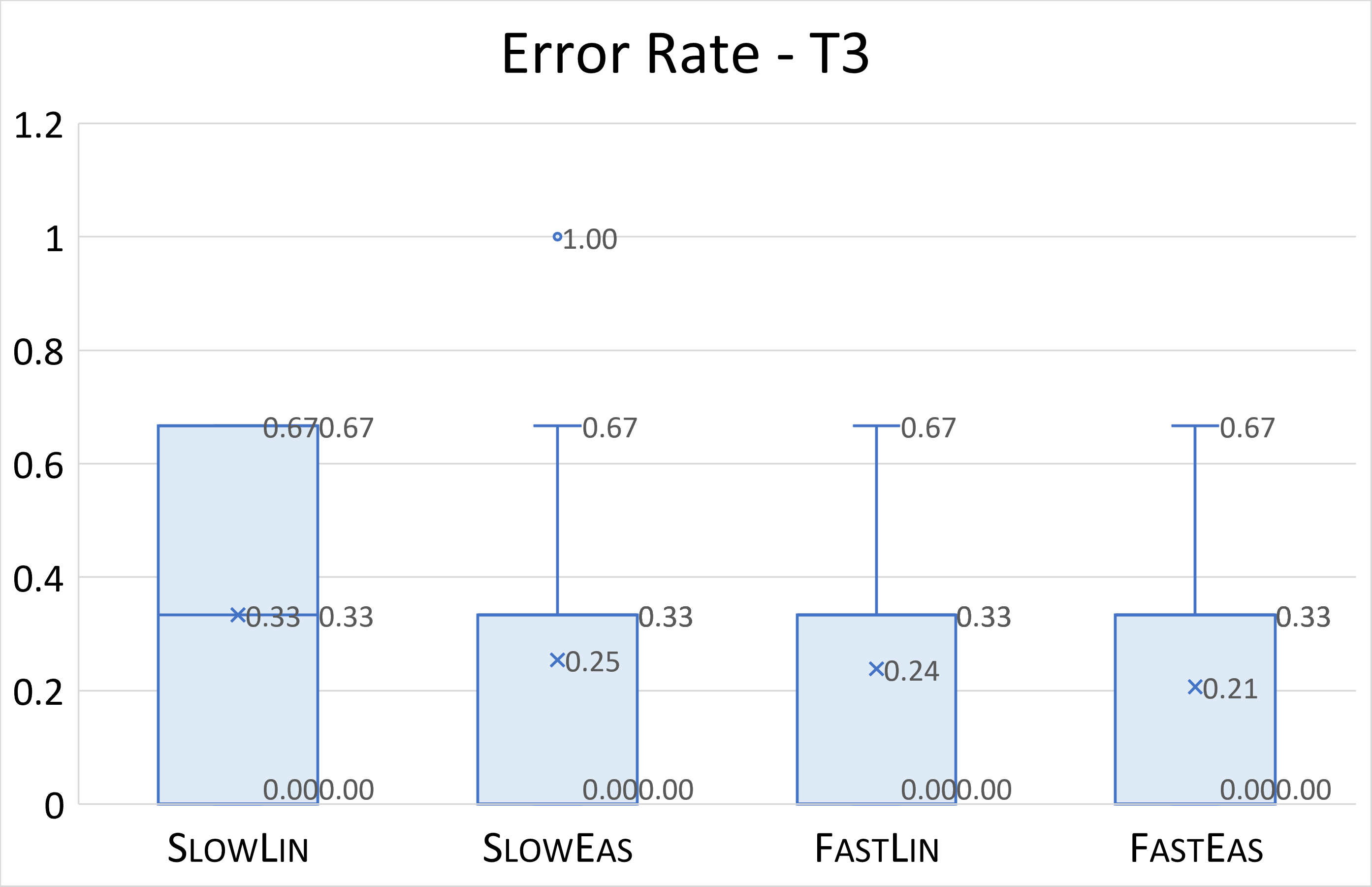}\label{fi:er-c}
    \hfil
    \includegraphics[width=.49\textwidth]{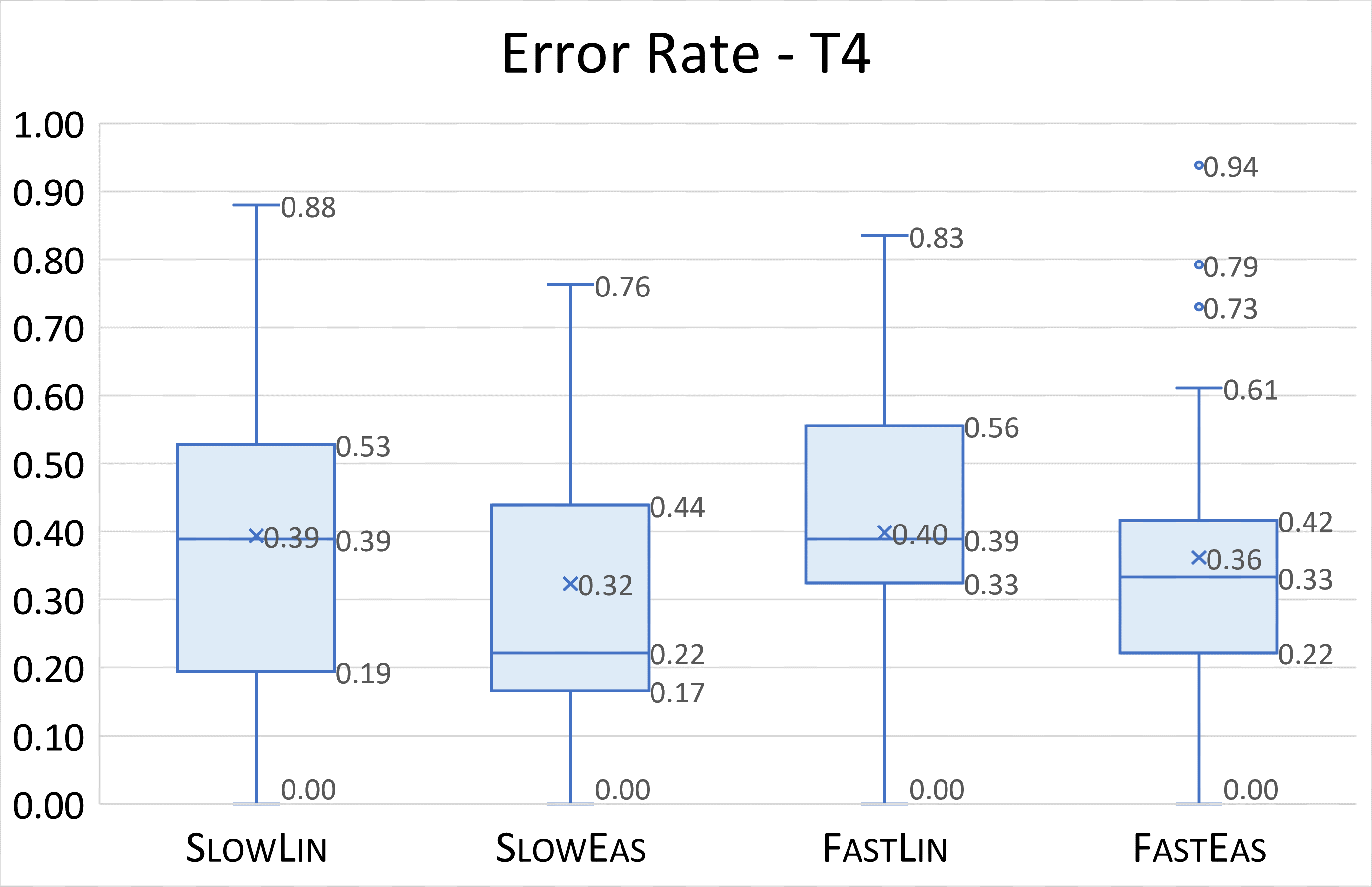}\label{fi:er-d}
    \hfil
    \includegraphics[width=.49\textwidth]{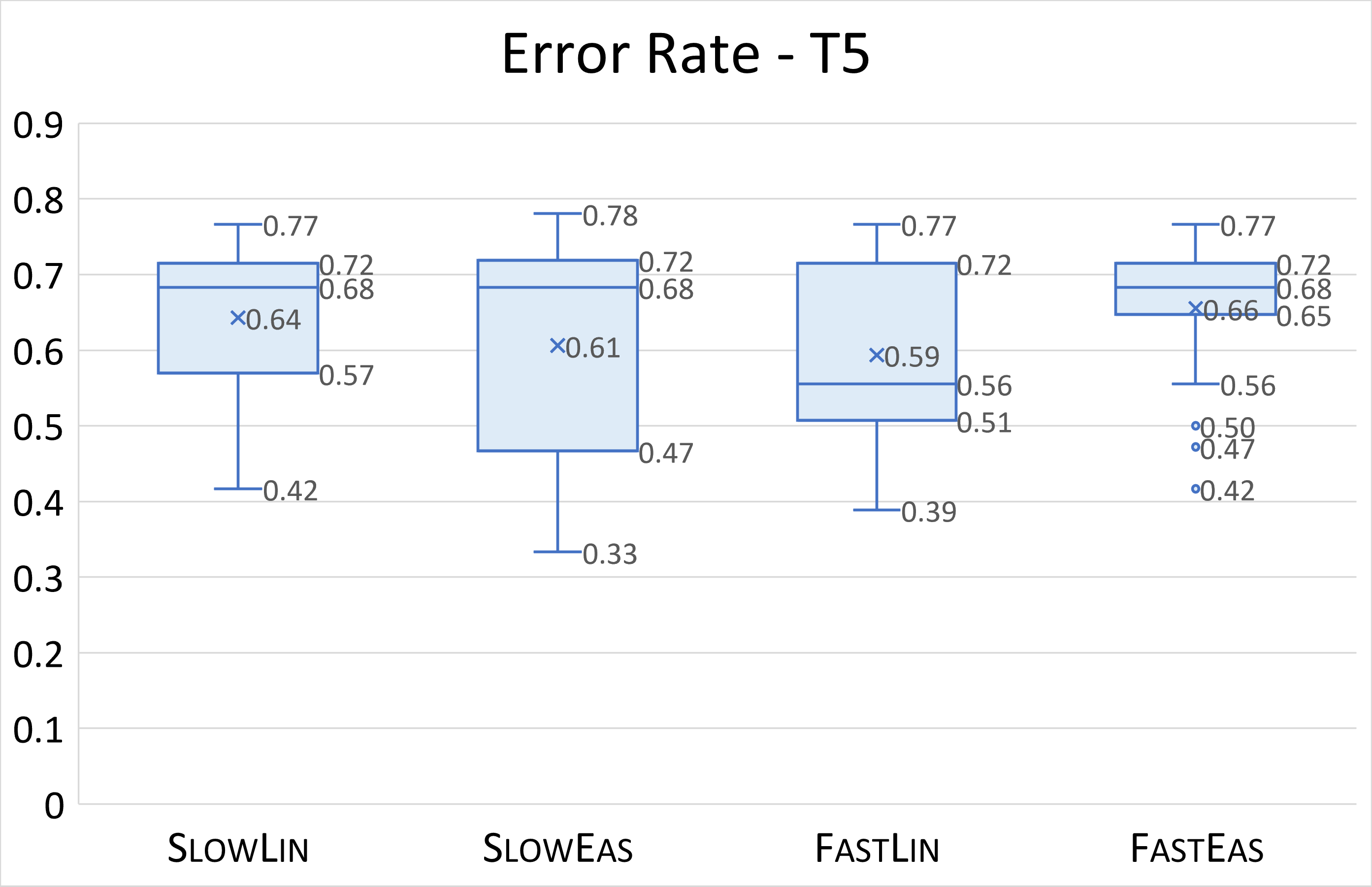}\label{fi:er-e}
    \hfil
    \includegraphics[width=.49\textwidth]{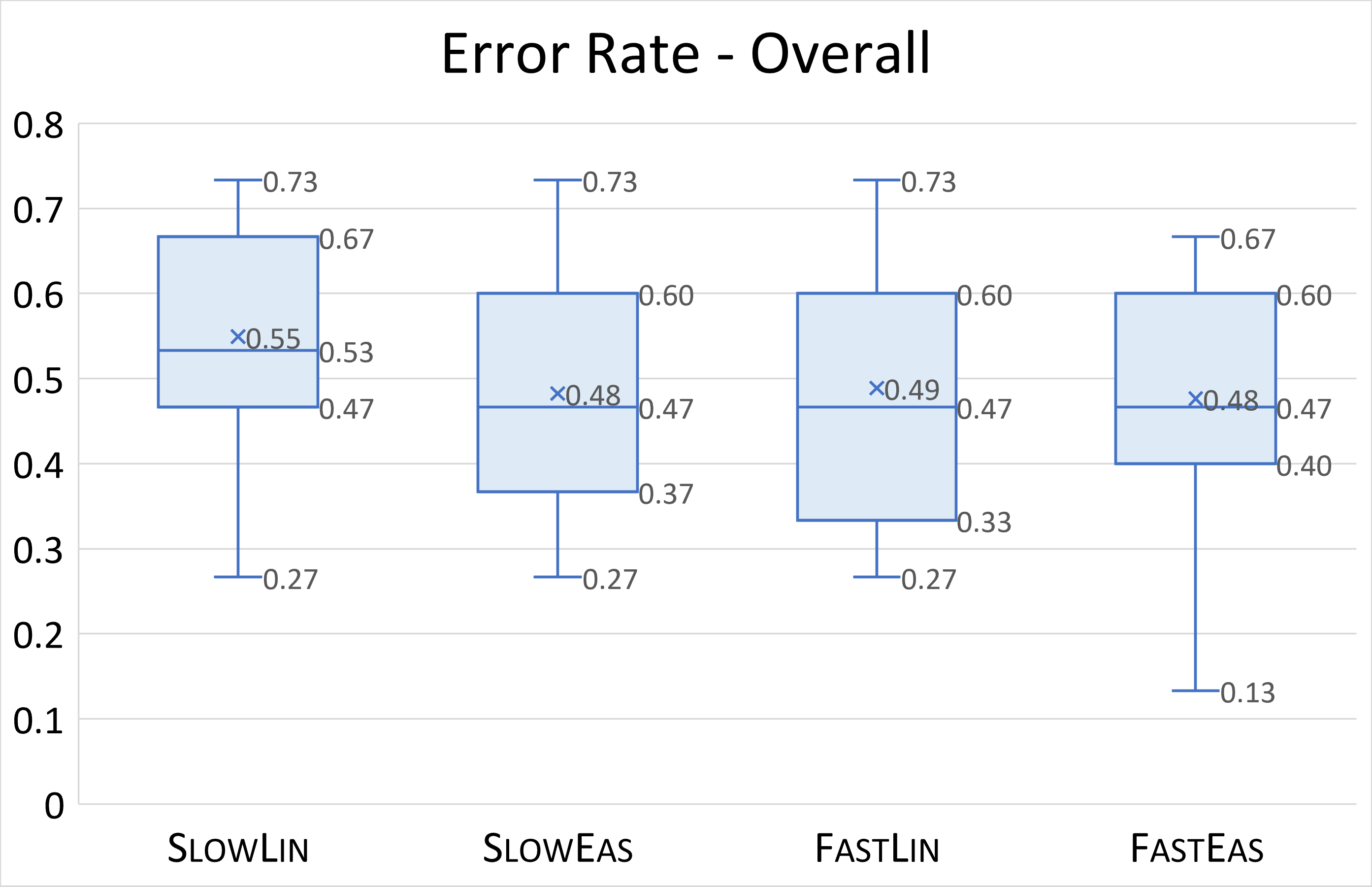}\label{fi:er-f}
    \caption{Error rate aggregated by task.}
    \label{fi:er-box}
\end{figure}

We performed a Shapiro-Wilk test~\cite{SHAPIRO01121965,thode-02} and we found that the populations were normally distributed for tasks \textsf{T2~(NeighborhoodSize)}, \textsf{T4~(CommonNeighbors)}, and \textsf{T5~(InterRegionEdges)}. We thus performed a one-way ANOVA test and post-hoc pairwise comparisons of the considered placement strategies by using Tukey’s HSD test~\cite{m2012}.
Regarding \textsf{T1~(Adjacency)} and \textsf{T3~(PathLength)}, we performed the non-parametric Kruskal-Wallis test. For all the above tests, the significance level was set to $\alpha = 0.05$. From the tests, we did not observe any statistically significant difference between the four experimental conditions. 
However, by looking at the box-plots of the response time and error rate, which are reported in Figs.~\ref{fi:rt-box} and~\ref{fi:er-box}, one can observe interesting phenomena\footnote{
Observe that the scales on the y-axes in Figs.~\ref{fi:rt-box} and~\ref{fi:er-box} differ between subfigures.
}.

Regarding response time, the plots in Fig.~\ref{fi:rt-box} show that for all tasks but~\textsf{T5~(InterRegionEdges)}, \texttt{Fast} speed and cubic Bézier curve \texttt{Easing} perform better than \texttt{Slow} speed and \texttt{Linear} easing, respectively. This behavior can also be observed over all tasks, but cubic Bézier curve \texttt{Easing} seems to perform better than \texttt{Linear} easing only if combined with \texttt{Fast} speed (\texttt{FastEas} model).
A similar trend is exhibited by the error rate reported in Fig.~\ref{fi:er-box}. In particular, for tasks \textsf{T2~(NeighborhoodSize)}, \textsf{T3~(PathLength)}, and \textsf{T4~(CommonNeighbors)} cubic Bézier curve \texttt{Easing} performs better than \texttt{Linear} easing in the case~of~equal~speed. Although average values are very low for \textsf{T1~(Adjacency)}, they show slightly better performance with \texttt{Fast} speed and cubic Bézier curve \texttt{Easing} than with \texttt{Slow} speed and \texttt{Linear} easing, respectively. This behavior can also be observed over all tasks. 

\subsection{Subjective Results}
The percentage of answers to the Likert scale questions are reported in Figs.~\ref{fi:qual-a} and~\ref{fi:qual-b} (the answer distributions as box-plots can be found in Appendix~\ref{app:charts}); we assigned a score from 1 (lowest) to
5 (highest) to each answer.
While there is no statistically significant difference among the conditions, we can observe the following.
    Regarding \textsf{beauty}, users seem to prefer \texttt{Slow} speed and cubic Bézier curve \texttt{Easing} rather than \texttt{Fast} speed and \texttt{Linear} easing. Indeed, the models with \texttt{Slow} speed received a percentage of positive appreciations (4 and 5) that is higher than the one obtained by the models with \texttt{Fast} speed, and the models with cubic Bézier curve \texttt{Easing} received a higher percentage of positive appreciations than the models with \texttt{Linear} easing. Also, the average ratings suggest that the preferred model is \texttt{SlowEas}.
    About \textsf{easiness}, on average \texttt{FastEas} and \texttt{SlowLin} are the models that were found to be most difficult and easiest, respectively.
    
    The percentage of users who considered the \textsf{speed} adequate is on average higher for the models with \texttt{Slow} speed than for the ones with \texttt{Fast} speed; see Fig.~\ref{fi:qual-c}.
    The percentage of users who found it \textsf{tiring} to work with \texttt{Fast} speed is higher than the one of users who worked with \texttt{Slow} speed; see Fig.~\ref{fi:qual-d}. Among the slow models, \texttt{Linear} easing seems to produce visualizations that are rated as less tiring than cubic Bézier curve \texttt{Easing}.

\begin{figure}[tb]
\centering
    \subfigure[]{\includegraphics[width=.49\textwidth]{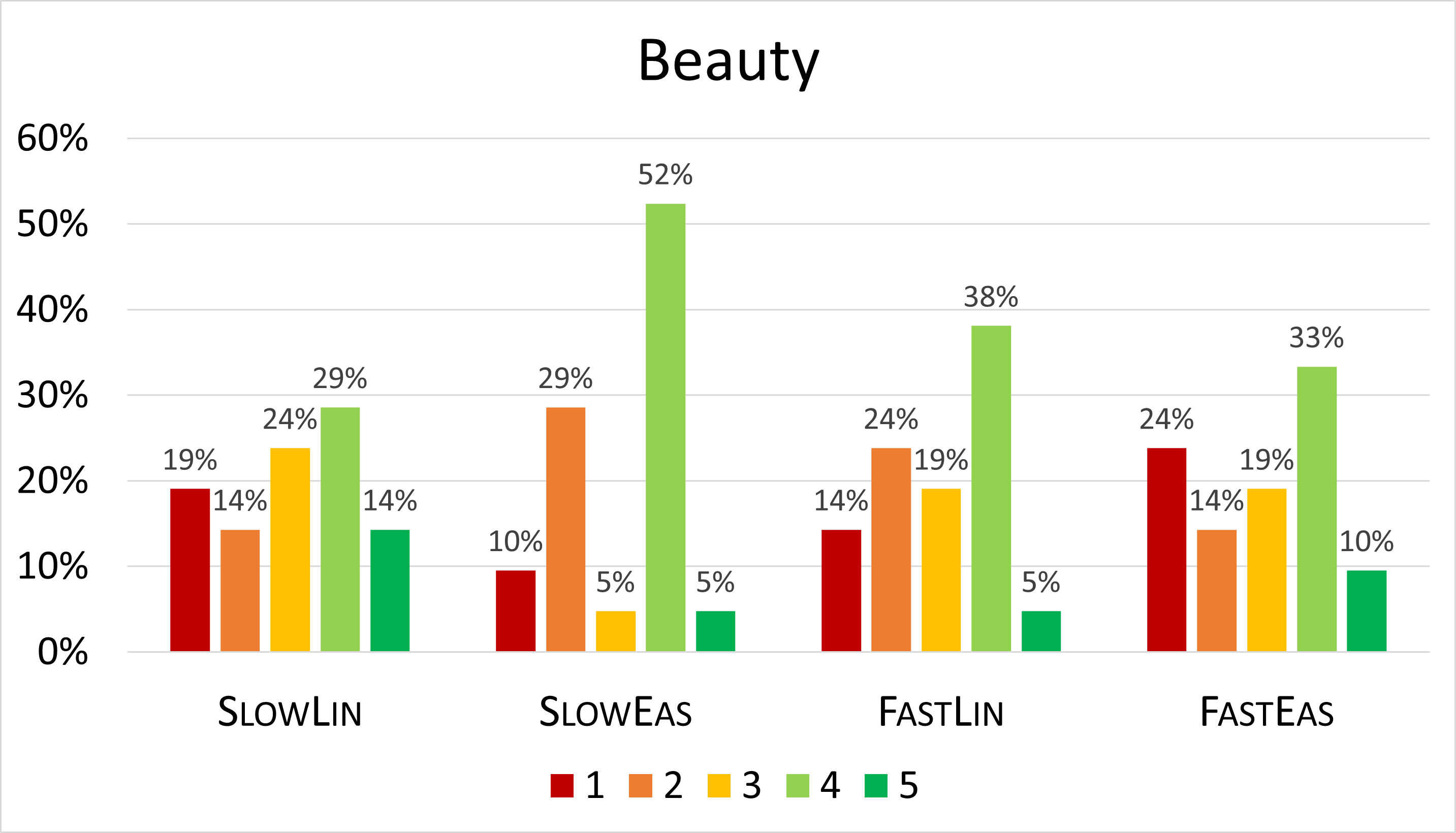}\label{fi:qual-a}}
    \hfil
    \subfigure[]{\includegraphics[width=.49\textwidth]{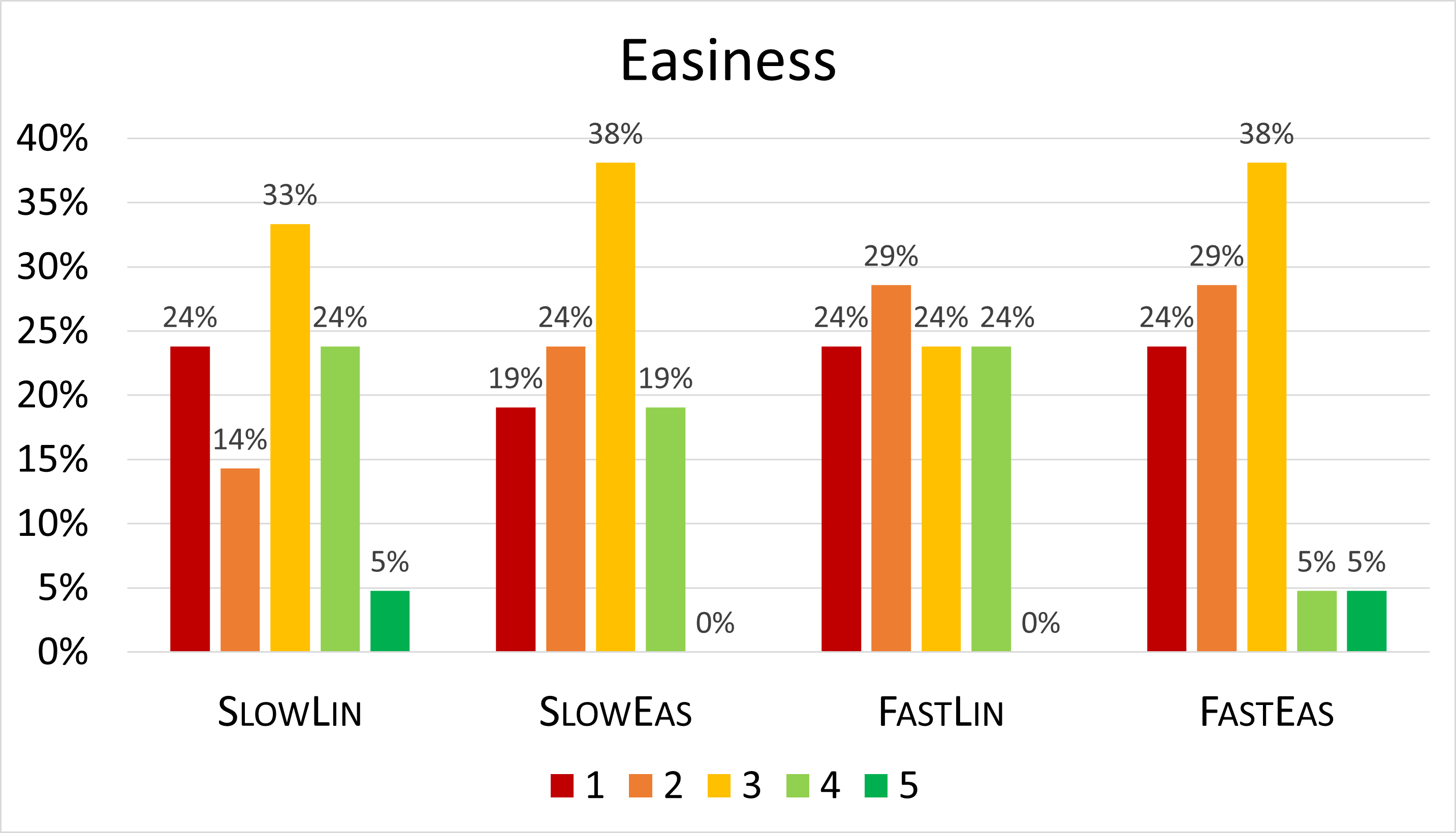}\label{fi:qual-b}}
    \hfil
    \subfigure[]{\includegraphics[width=.49\textwidth]{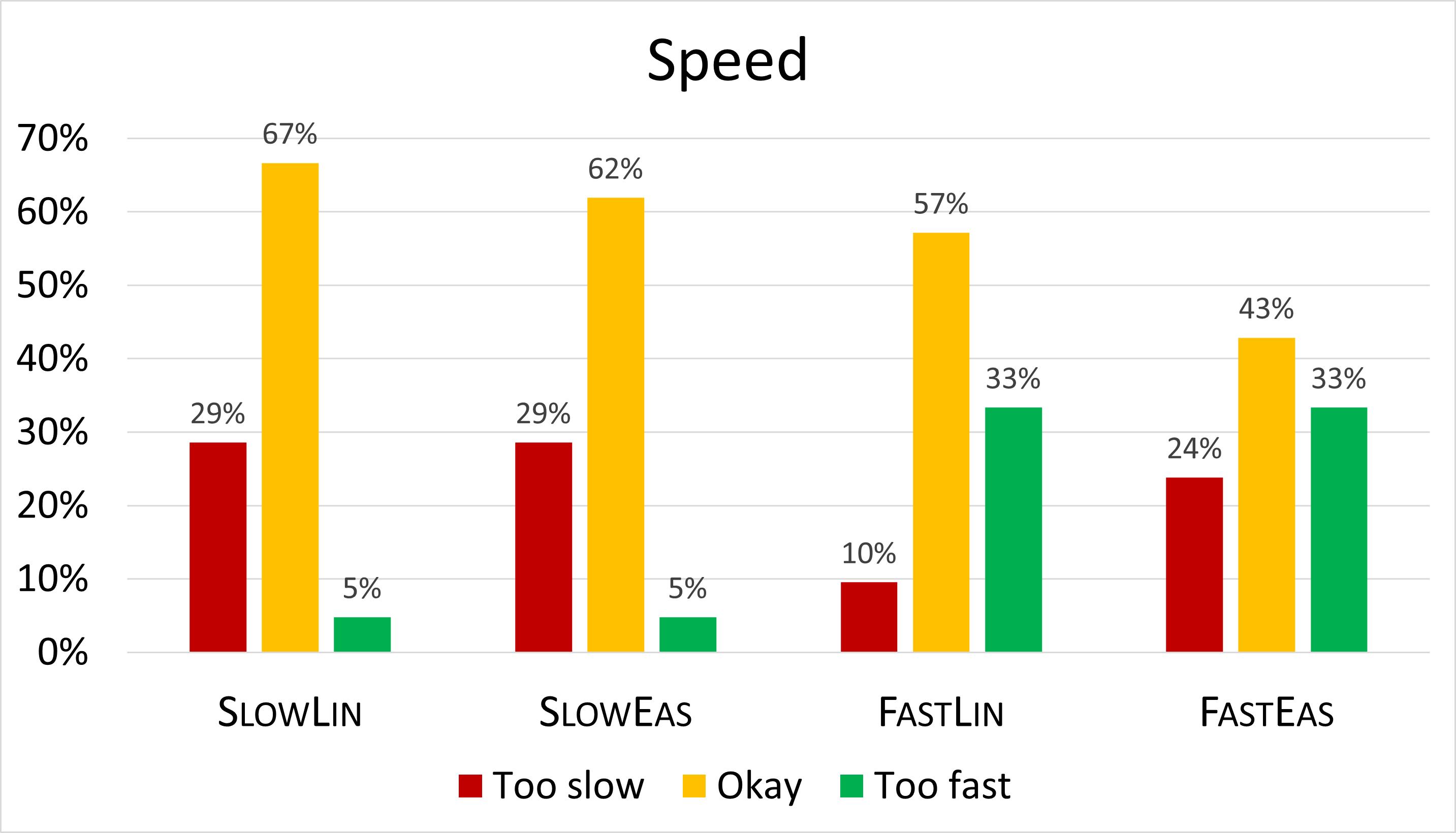}\label{fi:qual-c}}
    \hfil
    \subfigure[]{\includegraphics[width=.49\textwidth]{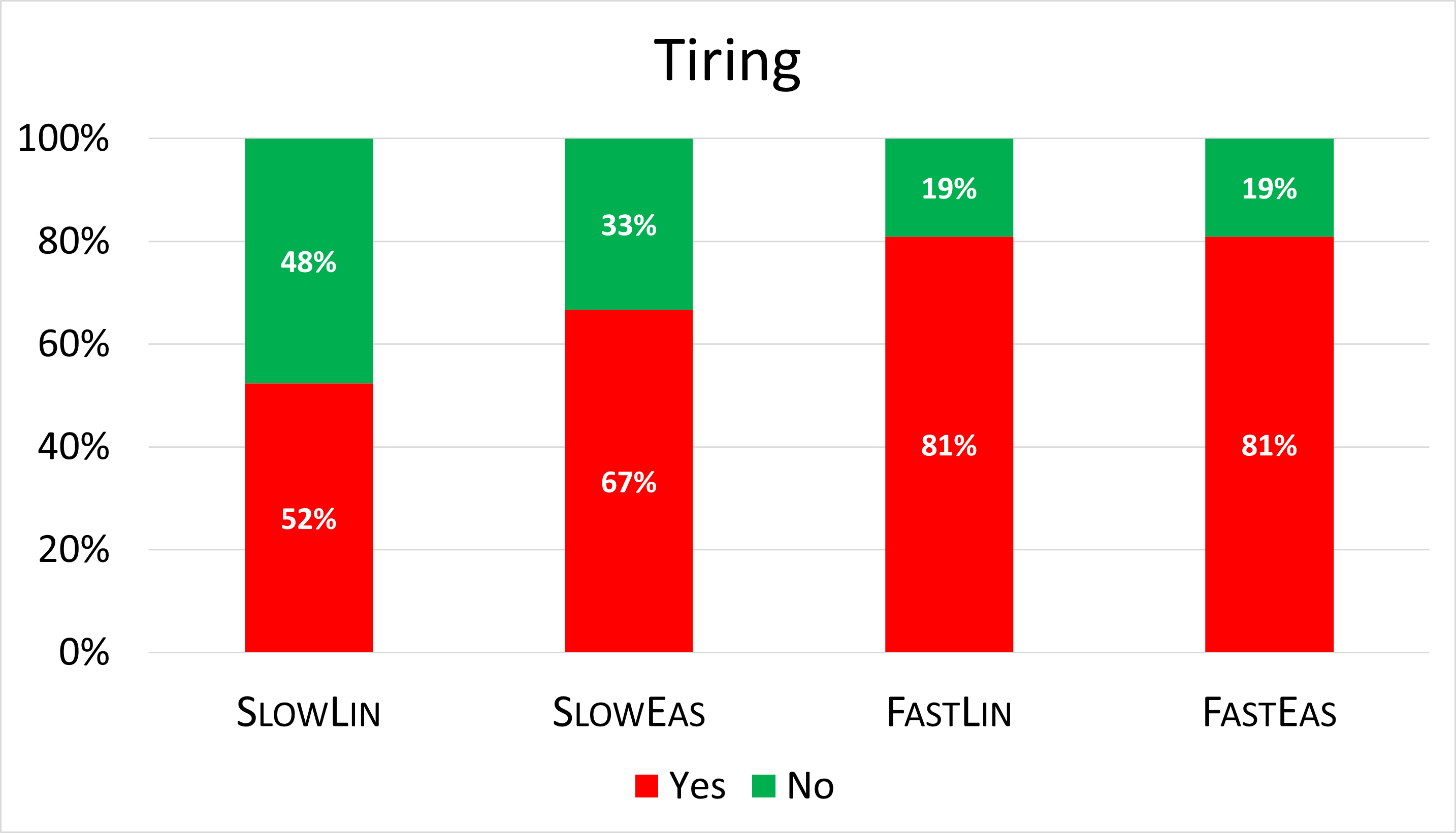}\label{fi:qual-d}}
    \caption{Subjective results. 
    In (a) and (b), 1=lowest and 5=highest.
    }
    \label{fi:qualitative}
\end{figure}

\begin{figure}[tb]
\centering
    \subfigure[]{\includegraphics[width=.49\textwidth]{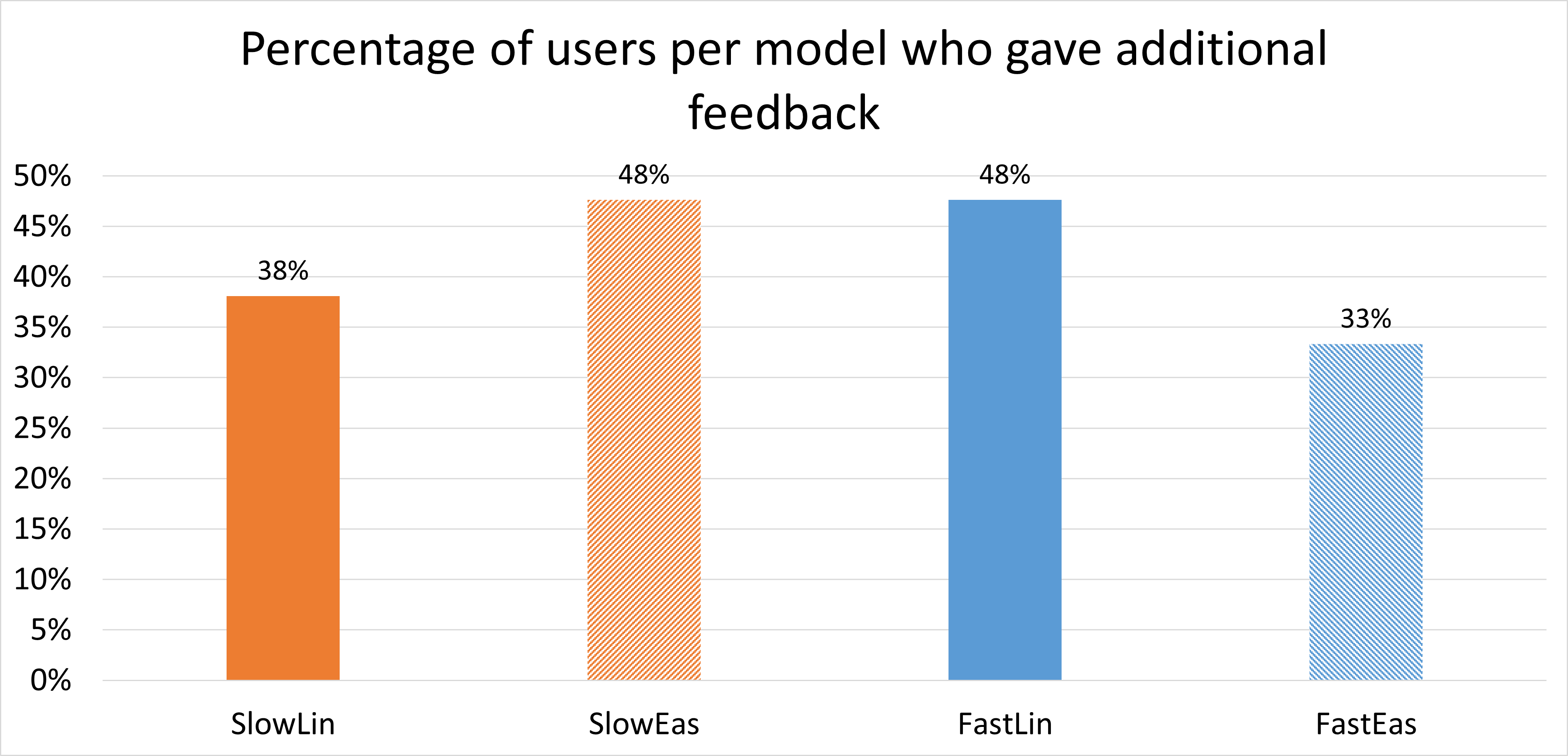}\label{fi:qual-percentage}}
    \hfil
    \subfigure[]{\includegraphics[width=.49\textwidth]{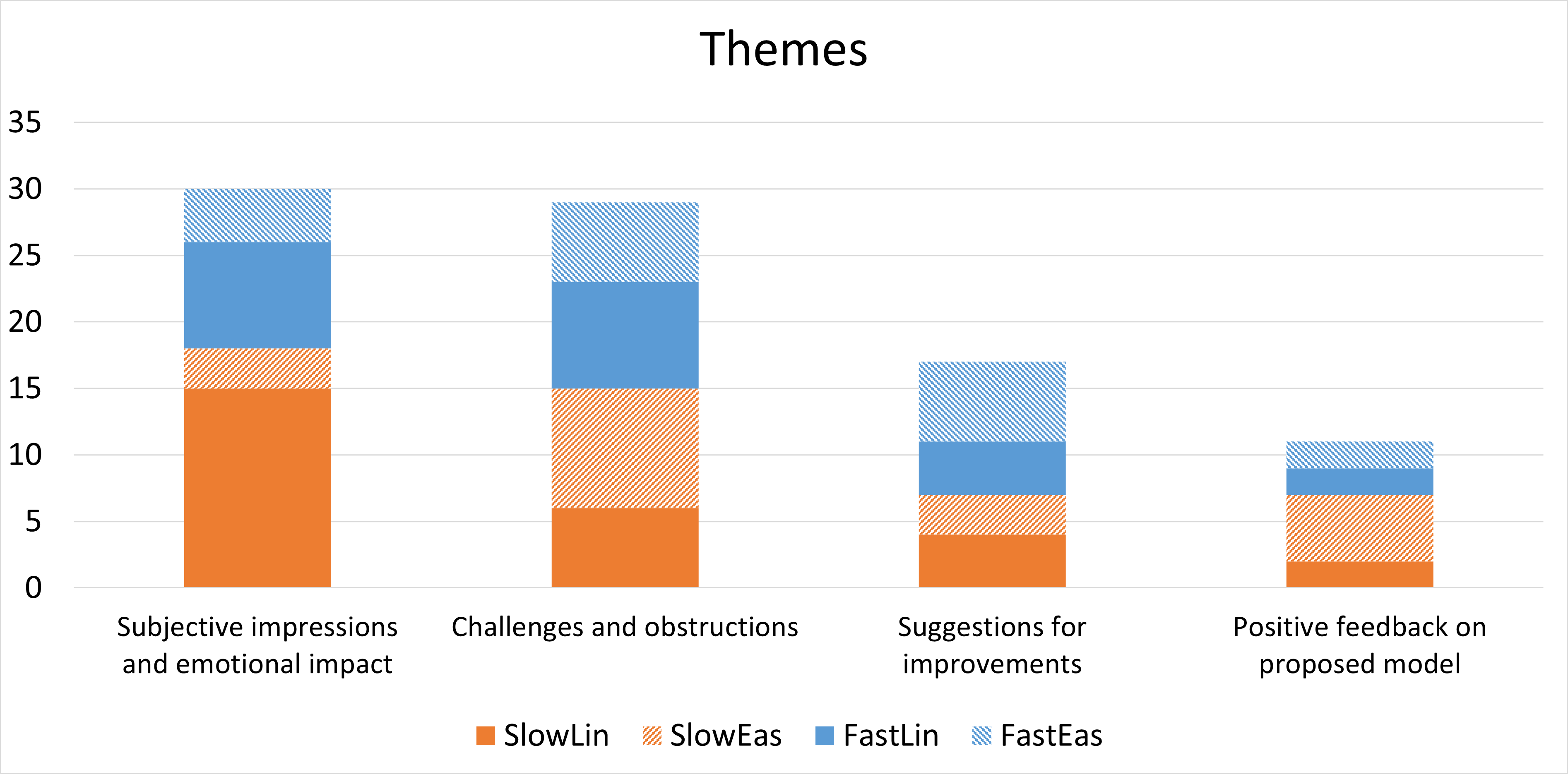}\label{fi:qual-themes}}
    \hfil
    \subfigure[]{\includegraphics[width=.49\textwidth]{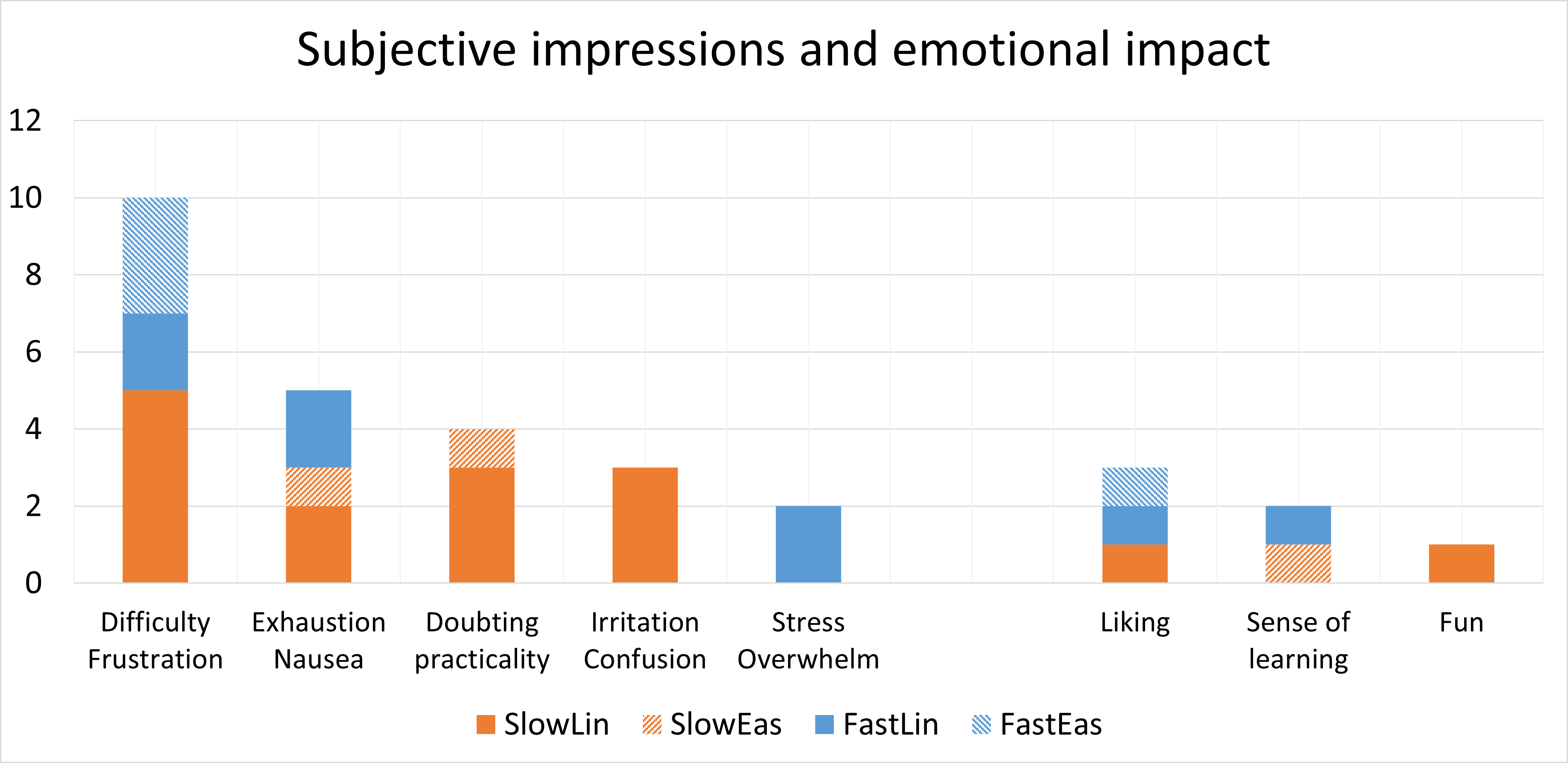}\label{fi:qual-emotions}}
    \hfil
    \subfigure[]{\includegraphics[width=.49\textwidth]{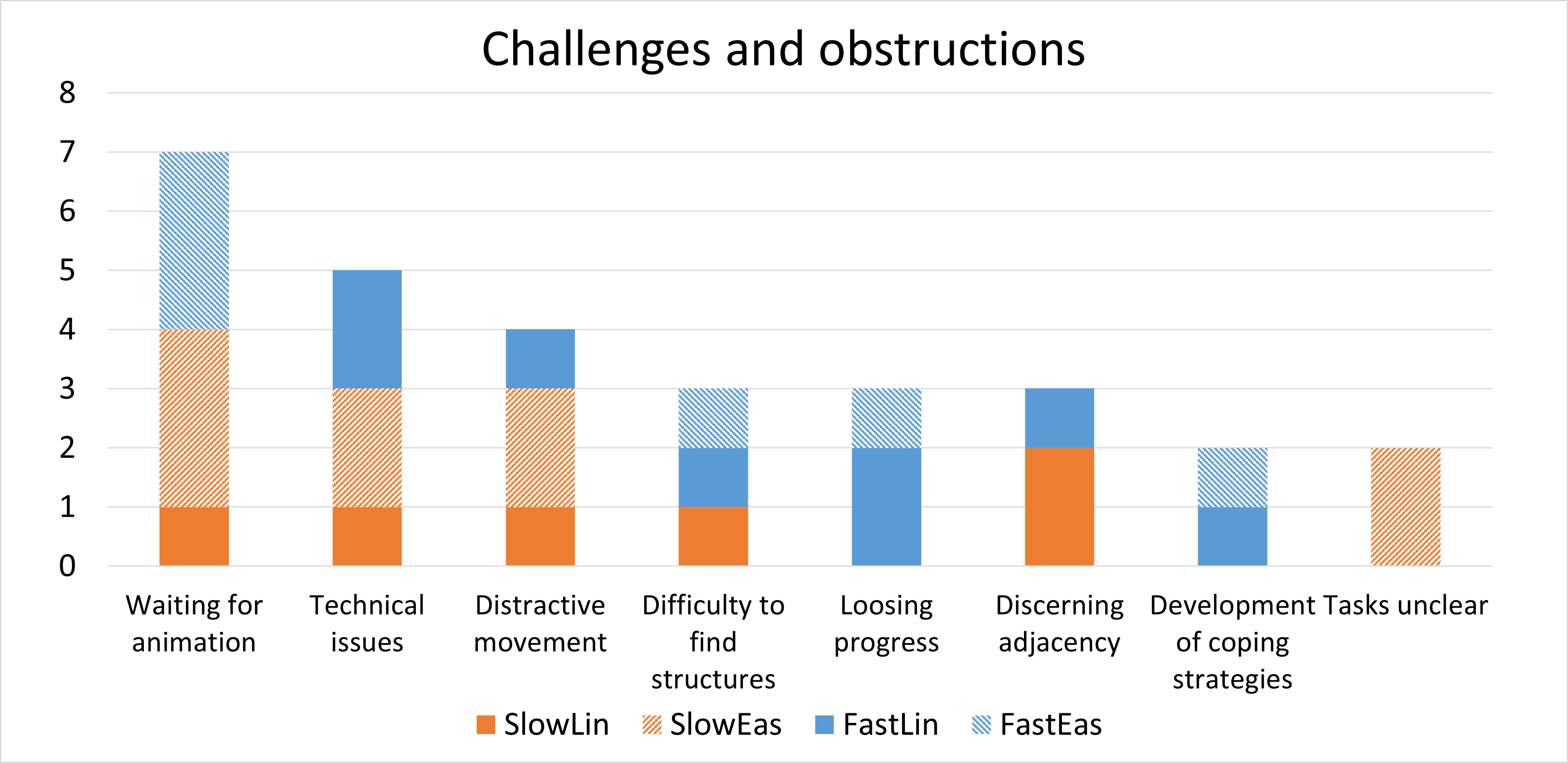}\label{fi:qual-challenges}}
    \hfil
    \subfigure[]{\includegraphics[width=.49\textwidth]{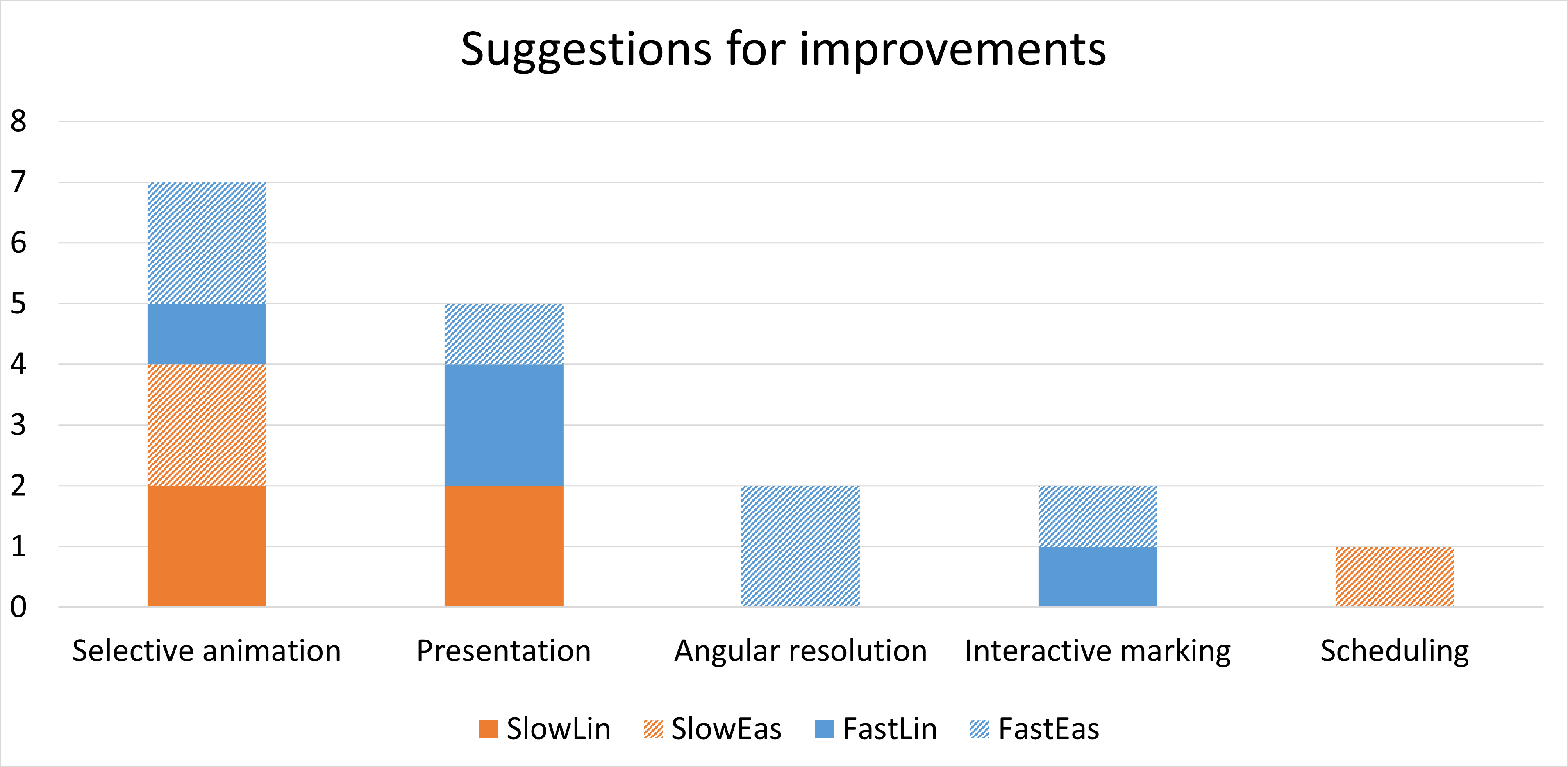}\label{fi:qual-improvement}}
    \hfil
    \subfigure[]{\includegraphics[width=.49\textwidth]{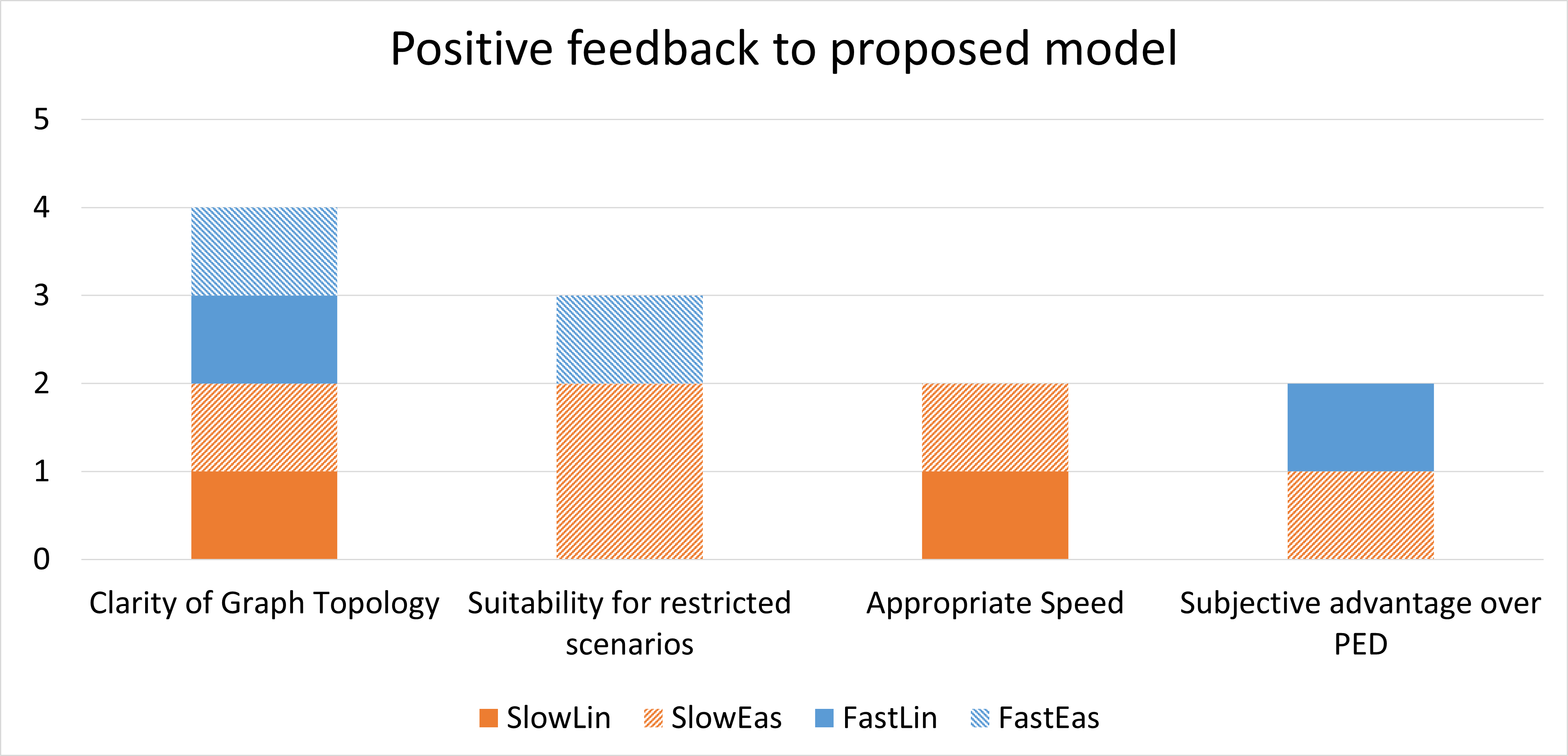}\label{fi:qual-positive-feedback}}
    \caption{Subjective results derived from additional user feedback.}
    \label{fi:qualitative-feedback}
\end{figure}

\smallskip
\noindent\textbf{Additional user feedback.}
Overall $42\%$ of users submitted additional feedback amounting to 35 submissions, while the proportion of submissions in each group was rather balanced, with \texttt{SlowEas} and \texttt{FastLin} giving the most comments ($10$ each); see Fig.~\ref{fi:qual-percentage}.
We performed a thematic analysis on the users' feedback and we identified four recurring themes, namely \emph{subjective impressions and emotional impact}~($30$), closely followed by \emph{challenges and obstructions} when performing the study~($29$), \emph{suggestions for improvements}~($17$), and \emph{positive feedback to the proposed model}~($11$); refer to Fig.~\ref{fi:qual-themes}. The different groups have largely made similar points, mostly with no clear trend to be noted. For a breakdown of each theme by model, see Figs.~\ref{fi:qual-emotions}--\ref{fi:qual-positive-feedback}.

\paragraph{Subjective impressions and emotional impact.}
Negative emotions~($5$ for \texttt{Easing}, $19$ for \texttt{Linear}) were reported more often than positive ones~($6$). The dominant factor was a sense of difficulty/frustration~($10$), followed by exhaustion/nausea~($5$), and doubting practicality~($4$, \texttt{Slow} only). Other negative sentiments are irritation/confusion ($3$, \texttt{Slow} only) and stress/overwhelm~($2$, \texttt{Fast} only). On the positive side, users reported a liking~($3$), a sense of learning~($2$), and fun~($1$).

\paragraph{Challenges and obstructions.} 
Users mostly described that waiting for the animation of an edge hindered them when performing the tasks~($7$, mostly \texttt{Easing}). Second most reported were technical issues~($5$, mostly delay or glitches in the animation), followed by a distraction due to the simultaneous movement~($4$). Other challenges were the general difficulty to find graph structures requested by the tasks~($3$), miscounting or losing progress~($3$, \texttt{Fast} only), discerning adjacency due to nodes placement~($3$), finding coping mechanisms to alleviate difficulty~($2$, \texttt{Fast} only) and not understanding task descriptions~($2$).

\paragraph{Suggestions for improvements.}
Users most frequently requested control over the animations ($7$, e.g., through hovering or clicking). Further, it was suggested to use more colors and different line-widths to highlight relevant structures~($5$). Concerning network layout, two users in the \texttt{FastEas} group suggested that a better angular resolution might improve task performance. The second requested interactive feature was the possibility to mark  nodes for tracking progress~($2$, \texttt{Fast} only). Lastly, there was one call for better animation scheduling. 

\paragraph{Positive feedback to proposed model.}
Most often it was mentioned that the animation made the recognition of the graph topology clearer~($4$), followed by the potential of the model in restricted scenarios, i.e., for certain graphs or tasks~($3$, \texttt{Easing} only). For the \texttt{Slow} speed, two users found the speed to be appropriate. Finally, two users experienced technical issues with animation: They continued solving the tasks as for \peds, which they described as much harder than~\meds.

\subsection{Discussion of our Results}
We did not observe statistically significant differences between the four models in terms of response time and error rate, which suggests that none of the models clearly outperforms the others in terms of readability.
However,
an analysis of the aggregate data provides some initial indications that cubic Bézier curve \texttt{Easing} may help users in executing topology-based tasks in an accurate manner.
We do not have any indications for the overview task (i.e., \textsf{T5~(InterRegionEdges)}), whose behavior seems not to be in line with the other tasks and the error rate is very high for all models (the percentages of wrong answers for the \texttt{marvel}, \texttt{boardgames},~and \texttt{kpop} networks are $100\%$, $99\%$, and $75\%$, respectively). 
Our interpretation is that~this task is not suitable for \meds, since the connections between the boundaries~of the highlighted parts of the network appear and disappear repeatedly, thus distracting the user and making it hard to keep track of the previously considered~edges.
Moreover, while we observed slightly lower response times for the \texttt{Fast} animation speed and the cubic Bézier curve \texttt{Easing} the collected data does not suggest a model clearly outperforming the others in terms of response time.

Referring to our hypotheses, the data
thus provide indications that only partially support \textsf{H1} and \textsf{H2}.
In particular, \texttt{Slow} speed does not give a lower error rate than \texttt{Fast} speed, while cubic Bézier curve \texttt{Easing} behaves better than \texttt{Linear} easing. Our interpretation for this phenomenon is that, when analyzing a connection between two nodes, the waiting time for the edge to appear might be long, and this is more evident with a slow animation. Further, the differences between \texttt{Linear} and cubic Bézier curve \texttt{Easing} may be caused by the circumstance that the longer time when stubs are drawn almost fully in cubic Bézier curve \texttt{Easing} helped the users when performing the tasks. 
Our subjective results support \textsf{H3}. The analysis of the users' free-form feedback suggests that the cubic Bézier curve \texttt{Easing} is preferred over \texttt{Linear} easing. 
Concerning the speed, the subjective feedback favors the \texttt{Slow} speed for \textsf{beauty}, \textsf{speed} appropriateness and \textsf{tiredness}, while the \textsf{easiness} of the tasks shows no clear trend. We believe that there are two different key effects at play here: First, the higher waiting times for the \texttt{Slow} speed are perceived by the users as a distracting factor as also mentioned in the free-form answers. Second, the \texttt{Fast} speed shows animations in different parts of the drawing in rapid succession, which may be overwhelming and lead to miscounting or losing progress. Thus, we think that an intermediate speed between 100 and 200 px/s would be most appropriate. This seems to be also partially supported by the answers about \textsf{speed}; see Fig.~\ref{fi:qual-c}. Also, note that \texttt{FastEas} has the highest maximum animation speed and the lowest ratings as an \emph{Okay} speed. 


\subsection{Limitations of our Experiment.}
%
    First, we did not observe statistically significant differences between the four models in terms of response time and error rate, which indicates that none of the models outperforms the others in terms of readability. Possible differences between the models could emerge with a higher number of participants or by considering a different study design, i.e., a within-subject experiment.  
    
    Second, we noticed that the only overview task that we considered (\textsf{T5~(InterRegionEdges)}) turned out to be very challenging for the users, which made us conclude that \meds are not a good paradigm to perform such a task. Perhaps, the choice of different overview tasks would have allowed us a better evaluation.
    
    Moreover, the choice of not allowing interaction implied to use networks that fit into a standard screen, thus facilitating the execution of an online test. We think that enabling interaction may allow the evaluation of our models on larger networks and could facilitate the users in the execution of tasks, e.g. by employing selective animation and marking. On the other hand, we believe that this would preferably require a  controlled experiment study design. 
    
    Finally, the readability of \meds may be sensitive to the scheduling algorithm used for the animation of edges and to the one used to produce the layout. This justifies further investigation with different scheduling and layout algorithms.

\section{Conclusions and Future Research Directions}\label{se:conclusion}

We presented a user study that evaluates the readability of \meds by comparing different speeds and easing functions for the animation of edges.
Since our results do not exhibit statistically significant differences from a task performance point of view, we suggest to follow user preferences. Namely,
%
users 
tend to favor the cubic Bézier curve \texttt{Easing} and rate slower animation speed as more beautiful and appropriate. Thus, choosing an intermediate animation speed may be the preferred option in practical applications. 

Our study has some limitations and cannot be generalized to settings significantly different from ours. This motivates
further experiments with larger networks, additional
tasks, different scheduling algorithms for the animation of edges, and interaction features.
Regarding the last aspect, we believe that the incorporation of user interaction can significantly improve the model by only showing animations of edges or nodes that are selected by the users. We emphasize that this has also been suggested by some participants of our experiment.

\bibliographystyle{splncs04}
\bibliography{refs}

\appendix
\newpage

\section{Algorithms for Computing MEDs}
\label{app:misueAkasaka}
Misue and Akasaka~\cite{DBLP:conf/gd/MisueA19} suggested a greedy algorithm to compute \meds of short total animation duration. Namely, they sorted edges in decreasing order of $\tau_{(u,v)}$ and started the animation of the next edge at the earliest time frame that do not introduce avoidable crossings. Misue~\cite{DBLP:conf/gd/Misue22} suggested three variants of the algorithm that aim to minimize the average time between two morphs of the same edge: 
\begin{enumerate*}
\item\label{variant:1} a variant that greedily picks edges to be animated again as long as there it is possible within $T$,
\item\label{variant:2} a variant that allows animations to begin before the start of $T$ and end after the end of $T$ yielding a cyclic overlap,
\item\label{variant:3} a variant that allows a certain number of avoidable crossings.
\end{enumerate*}
For the purpose of this study we decided to implement Variant~\ref{variant:1} and ignore the latter two variants. Namely, Variant~\ref{variant:2} removes the initial time frame during which all edges are at stub length ratio $\delta_0$, which may serve as a useful anchor point for a viewer, while only yielding minor improvements~\cite{DBLP:conf/gd/Misue22}. Variant~\ref{variant:3} introduces the number of crossings as an additional variable that may interfere with our experiments.
While the algorithmic descriptions of Misue and Akasaka~\cite{DBLP:conf/gd/Misue22,DBLP:conf/gd/MisueA19} do not include $\tau_{1/2}$ and $\tau_{distinct}$, it is straight-forward to add them. Meanwhile, the algorithms work for any invertible function $\delta_{(u,v)}$ which for our purposes require that $\eta$ is invertible.

\section{Box-plots for Subjective Analysis}\label{app:charts}
\begin{figure}[h]
\centering
    \includegraphics[width=.49\textwidth]{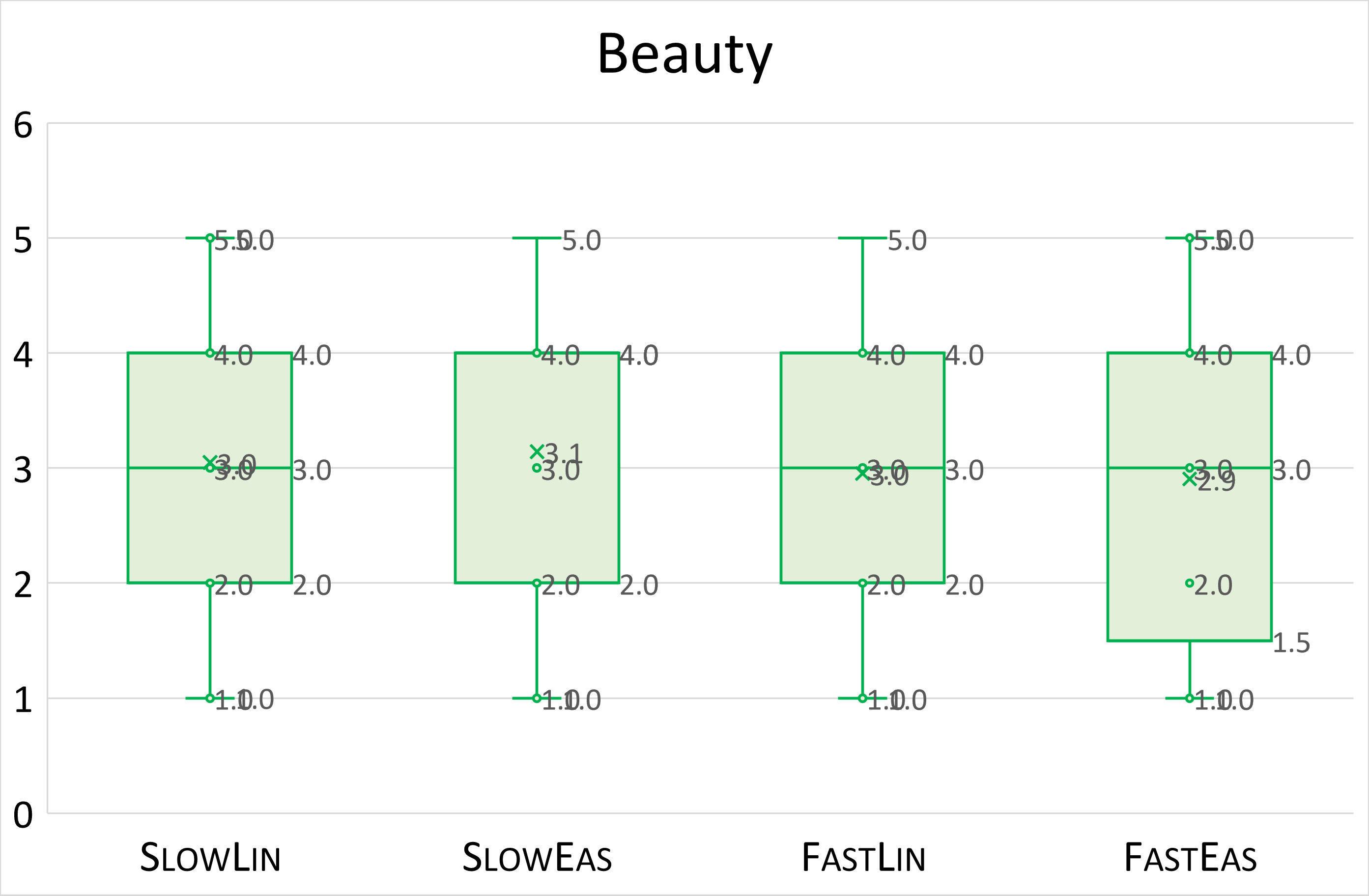}\label{fi:qual-box-a}
    \hfil
    \includegraphics[width=.49\textwidth]{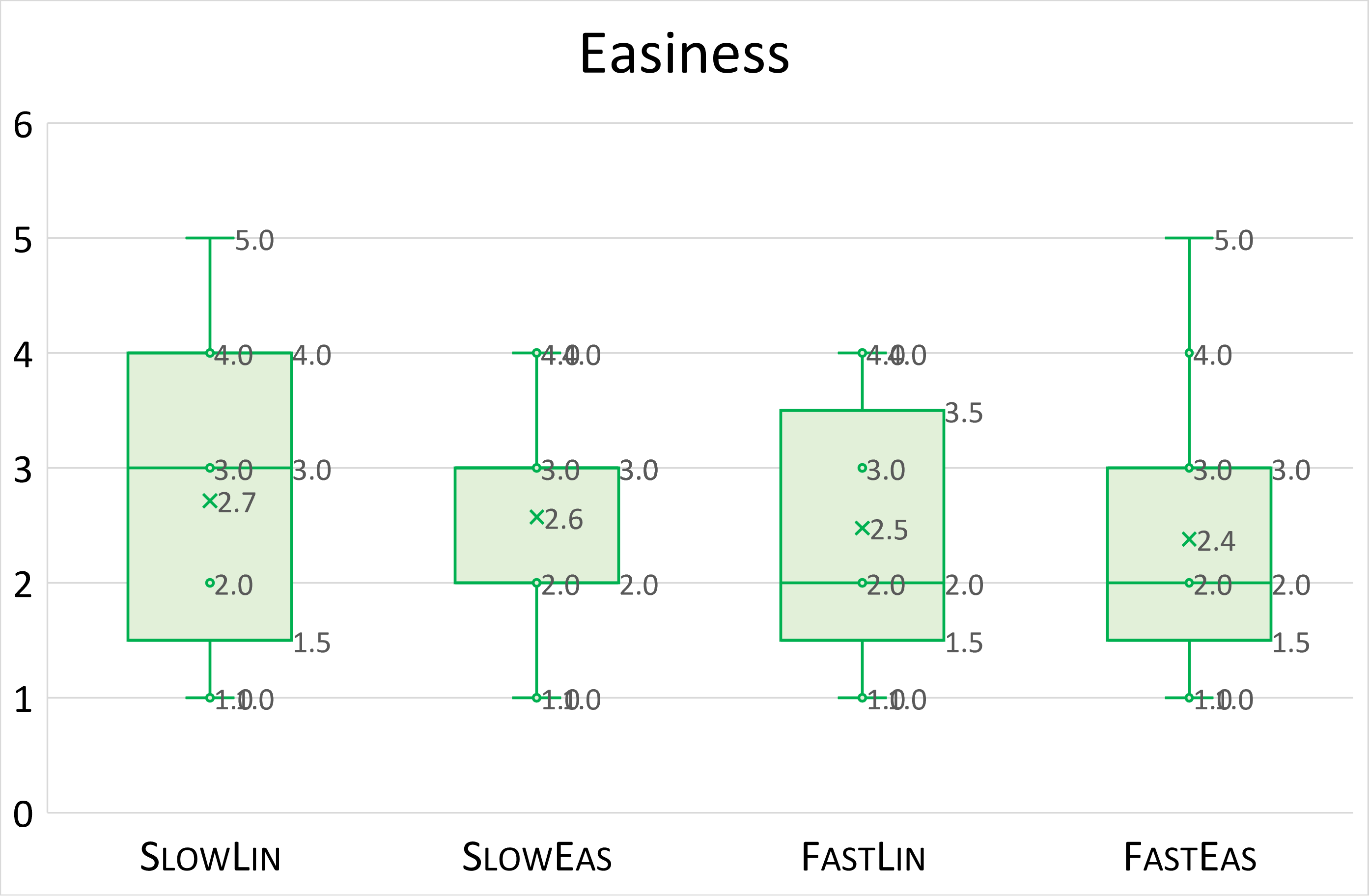}\label{fi:qual-box-b}
    \caption{Box-plots concerning subjective results.}
    \label{fi:qualitative-box}
\end{figure}
\end{document}